# Semi-Implicit Stellarator Magnetohydrodynamics with Nodal Spectral Elements

**University of Wisconsin-Madison**
**Center for Plasma Theory and Computation**
**Report UW CPTC 25-1**

Carl Sovinec,[1] Sanket Patil,[2] and Vincent Cauilan[1]
[1]Department of Nuclear Engineering & Engineering Physics
[2]Department of Physics
University of Wisconsin-Madison
Madison, WI 53706-1609

November 15, 2025

# Semi-Implicit Stellarator Magnetohydrodynamics with Nodal Spectral Elements

C. R. Sovinec,[1] S. A. Patil,[2] and J. V. Cauilan[1]

*University of Wisconsin-Madison, [1]Dept. of Nuclear Engineering & Engineering Physics and [2]Dept. of Physics*

Nonlinear time-dependent computation of macroscale dynamics in stellarators is motivated by laboratory results showing the possibility of robust operation in conditions where magnetohydrodynamic (MHD) modes are linearly unstable. A new formulation of semi-implicit MHD computation for toroidally shaped magnetic confinement systems uses 2D nodal spectral elements over the poloidal plane and Fourier representation over a generalized toroidal angle. Geometric mappings and steady-state (equilibrium) fields are expanded in the same 3D representation as the time-evolved fields to model non-axisymmetric configurations. For accuracy at large timestep, the semi-implicit operator is based on the ideal-MHD energy integral using 3D pressure and magnetic fields. The nodal spectral elements allow numerical convergence through either *h*-refinement or *p*-refinement. Our implementation (NIMSTELL) with the continuous *H*1 expansion of magnetic-field components and diffusive divergence control is a generalization of the NIMROD code [JCOMP **195**, 355 (2004)]. The NIMSTELL implementation is verified linearly and nonlinearly on resonant ideal interchange, where convergence from the stable side results from the stabilization method used in NIMROD [JCOMP **319**, 61 (2016)]. Optionally, NIMSTELL may use an ***H***(curl) representation for vector potential, and both magnetic representations are verified with respect to results from JOREK [Phys. Plasmas **29**, 063901 (2022)] on linear and nonlinear magnetic tearing in the W7-A rotating-ellipse configuration. Application of the existing vector-potential implementation to interchange shows that it needs a minimum level of electrical resistivity to avoid numerical noise for a given level of spatial resolution. Solving the algebraic systems from the implicit parts of the time advance is facilitated by including the Fourier components of stellarator mode families in each preconditioning operation.

## 1. Introduction

One of the many important applications of numerical magnetohydrodynamics (MHD) computation is the study of macroscale dynamics in magnetically confined plasma. The MHD model simplifies significant aspects of high-temperature plasma dynamics, but it provides critical insight on instabilities driven by spatial inhomogeneity and the consequences of instability. Moreover, different levels of MHD computation satisfy a variety of physical research needs. Computations of potential energy in the linearized ideal MHD system address stability at minimal cost [1]. Eigenmode computations for linear ideal [2] and nonideal [3], [4] behavior produce the spectra of frequencies and growth rates and eigenfunctions. In MHD's complete form, time-dependent nonlinear computations are used to study whether instabilities saturate in relatively benign topology changes [5], [6], relax the configuration to more stable configurations [7], [8], or lead to disruptive evolution [9], [10], [11], [12].

The majority of MHD computations for magnetic confinement to date have been applied to configurations where electrical current flowing in the plasma provides the necessary rotational transform, *i.e.*, twist, of the magnetic field to confine charged particles in toroidal configurations.

Stellarators and torsatrons are toroidally asymmetric configurations, where the magnetic twist results from shaped external coils. This approach to magnetic confinement avoids free energy from spatial variation of current density running along magnetic field in the plasma [13]. Nonetheless, macroscale dynamics driven by plasma pressure and residual plasma current can be important. Here, we present development for time-dependent simulations of stellarators and torsatrons.

An open research question that motivates MHD computation for stellarators is what limits the plasma pressure for a given magnetic field. Unlike the symmetric tokamak configuration that relies on plasma current, nonlinear evolution from linear instability often remains non-disruptive in stellarators. The largest attained ratio of average plasma pressure and magnetic pressure, $\langle\beta\rangle$, of more than 4.5% was achieved in the Large Helical Device (LHD), whose design was not optimized for MHD [14]. The helical coil currents were distributed to produce a large plasma aspect ratio, reducing the Shafranov shift that moves the central part of the plasma toward the outboard side of the torus. The high-$\langle\beta\rangle$ LHD discharges exceed the Mercier linear ideal-MHD stability limit [15] with seemingly little effect, possibly due to LHD's relatively large magnetic shear [16]. Relaxation-type computations with the HINT code show loss of outer flux surfaces with increasing Shafranov shift [17], [18]. Similarly, the Wendelstein 7-AS experiment achieved MHD-quiescent operation at $\langle\beta\rangle = 3.4\%$ [17]. Appreciable Shafranov shift was observed in its high-$\langle\beta\rangle$ operation, and relaxation computations with the PIES code [19] indicate chaotic field-line topology over the outer part of the plasma [17]. Earlier operation in W7-AS included an initial inward-shifted configuration to allow large Shafranov shift. The magnetic well property, which enhances linear MHD stability [13], is reduced with the inward shift, but the experiment only produced non-disruptive MHD relaxation events that lost approximately 10% of the stored thermal energy [20].

Time-dependent MHD computations for stellarators can help address this and other important questions. Early contributions include computation of two-fluid effects on ideal and resistive ballooning using the finite-element version of the original M3D code [21]. A study of partial pressure-profile collapses in inward-shifted LHD discharges iterated between the resistive reduced-MHD dynamical evolution with the NORM code and updates of the equilibrium profile [6] from VMEC [22]. The role of nonlinear resistive ballooning in LHD core density collapses was investigated with the explicit MIPS code [23]. More recent time-dependent stellarator MHD codes are redevelopments—to varying extent—of existing macroscale simulation codes for toroidally symmetric configurations. The stellarator version of the M3D-C1 code [24] applies its 3D finite-element representation to vary the shape of the poloidal plane over the toroidal angle [25]. It retains the potential representation of the M3D equations for full MHD evolution. It has been applied to pressure-induced MHD [26] and magnetic reconnection events associated with electron cyclotron current drive [27] in the W-7X experiment and to interchange dynamics in an optimized quasi-axisymmetric (QA) configuration [28]. Generalizing JOREK [29], [30] required theoretical development of its "ansatz based" reduced MHD [31]. The implementation retains JOREK's Fourier representation over the toroidal angle and applies it to geometric information in addition to the physical fields [32]. Recent work adds parallel flow evolution to the implementation and extends benchmarks and validation to the W7-AS configuration [33]. Both M3D-C1 and JOREK use implicit time-stepping to avoid numerical stability limitations.

Analogous to these developments for M3D-C1 and JOREK, the implementation described here, NIMSTELL, is a stellarator variant of the full-MHD NIMROD code [34], which has been applied to tokamaks, reversed-field pinches, spheromaks, and other magnetic confinement and space-plasma configurations. The NIMROD implementation assumes that the geometry and prescribed steady-state fields are symmetric with respect to a periodic coordinate, which is typically taken as the toroidal angle. Previous application of NIMROD to stellarator configurations include the study of pressure-driven topology evolution in a straight stellarator [35], evolution of the rotational-transform profile subject to Ohmic current drive in the Compact Toroidal Hybrid (CTH) experiment [36], and flux-surface loss and chaotic-field transport in a 10-field period torsatron [37]. A study of sawtoothing from Ohmic current drive in CTH investigates the influence of external transform on current-driven dynamics and quantifies the numerical cost of representing stellarator fields with an axisymmetric mesh [38]. These applications satisfy NIMROD's geometric limitation, but many modern stellarator experiments are not tractable with NIMROD. Besides computational efficiency, rigid external coils cannot be represented within NIMROD's computational domain. The coil fields must be represented through magnetic-field boundary conditions along a symmetric surface that lies between the plasma surface and the physical coils, and such surfaces do not exist in modern experiments and designs. The development of NIMSTELL alleviates these restrictions through Fourier expansions of geometry and steady-state fields over a generalized toroidal coordinate.

With respect to the other recent stellarator MHD implementations, NIMSTELL adds a unique combination of numerical methods that are described in this article. NIMSTELL solves the non-reduced MHD system and has spectral spatial representation capability with its 2D nodal spectral elements [39], [40] combined with Fourier series for the periodic coordinate. It also uses a semi-implicit time-advance of primitive variables for flow-velocity and magnetic field, though vector potential is an option for NIMSTELL. For comparison, M3D-C1 solves non-reduced systems with potential fields for flow-velocity and for magnetic field and uses reduced quintic triangles for the poloidal plane and Hermite cubics over the toroidal angle [25], [41]. Having continuous first derivatives in its spatial representation allows differentiation up to fourth order for its potential formulation. JOREK is typically used to solve reduced MHD systems. It uses 2D Bezier elements that can have high-order geometric continuity for the poloidal plane and 1D Fourier series for the toroidal angle [30], [42]. Both M3D-C1 and JOREK use implicit time advances [43], [30]. Their consistent implicit formulations allow accuracy at larger time-step than NIMROD's semi-implicit methods, but there is a tradeoff in the size and cost of the algebraic systems that need to be solve at each step [44]. The leapfrog-based semi-implicit method of NIMROD and NIMSTELL solves one physical field at a time. We expect this property to be helpful for stellarator applications, as it is for others, in reducing the size of matrices used for preconditioning.

The remainder of this article is organized as follows: The MHD modeling is described in Section 2. Numerical methods are presented in Section 3. Test applications are presented in Section 4, and conclusion are discussed in Section 5. Additional topics related to NIMSTELL's vector-potential option are covered in the appendices.

## 2. Mathematical model

In this article, we focus on solving the time-dependent equations of MHD. Our model is non-ideal in the sense that there is resistive electric field in Ohm's law and viscous dissipation in the plasma fluid-velocity equation. For magnetic confinement, we include the important extensions of anisotropic thermal conduction and viscous stress, where the orientation of the anisotropy follows the magnetic field.

The nonlinear evolution equations for the plasma-fluid model solved by NIMSTELL are:

$$\frac{\partial n}{\partial t} + \nabla \cdot (n\boldsymbol{V} - D_n \nabla n) = 0 \,, \tag{1}$$

$$\rho\left(\frac{\partial}{\partial t} + \boldsymbol{V}\cdot\nabla\right)\boldsymbol{V} = \nabla \cdot \left[\frac{1}{\mu_0}\boldsymbol{BB} - \underline{\mathbf{I}}\left(\frac{B^2}{2\mu_0} + p\right) - \underline{\boldsymbol{\Pi}}\right], \tag{2}$$

$$\frac{nk_B}{\Gamma-1}\left(\frac{\partial}{\partial t}+\boldsymbol{V}\cdot\nabla\right)T=-nk_BT\nabla\cdot\boldsymbol{V}-\nabla\cdot\boldsymbol{q}+Q\,. \tag{3}$$

Mass density $\rho$ and electron particle density $n$ are related by $\rho = mn$ with $m$ being the effective ion mass for a given charge state. When the charge state is 1 and electrons and ions have the same temperature $T$, plasma pressure is $p = 2nk_BT$. The $\Gamma$-factor in Eq. (3) is the adiabatic index, and $D_n$ is a numerical particle diffusivity that helps avoid numerical noise.

This single-fluid, single-temperature model is closed by the stress tensor $\underline{\boldsymbol{\Pi}}$ and the heat-flux density $\boldsymbol{q}$, and $Q$ represents a heat-source density. The conductive heat-flux closure for Eq. (3),

$$\boldsymbol{q}=-(\kappa_\parallel-\kappa_{\text{iso}})\hat{\boldsymbol{b}}\hat{\boldsymbol{b}}\cdot\nabla k_BT-\kappa_{\text{iso}}\nabla k_BT\,, \tag{4}$$

is a computationally tractable approximation for high-temperature magnetized plasma that represents anisotropy though distinct parallel $\kappa_\parallel$ and perpendicular $\kappa_\perp$ thermal conductivities with respect to the magnetic direction vector $\hat{\boldsymbol{b}} = \boldsymbol{B}/B$. Similarly, a viscous closure for Eq. (2) uses

$$\underline{\boldsymbol{\Pi}}=-\rho\left[\frac{1}{2}\nu_\parallel\hat{\boldsymbol{b}}\cdot\underline{\boldsymbol{W}}\cdot\hat{\boldsymbol{b}}\left(3\hat{\boldsymbol{b}}\hat{\boldsymbol{b}}-\underline{\boldsymbol{I}}\right)+\nu_{\text{iso}}\underline{\boldsymbol{W}}\right], \tag{5}$$

where $\underline{\boldsymbol{W}} = \nabla\boldsymbol{V} + \nabla\boldsymbol{V}^T - (2/3)\underline{\boldsymbol{I}}\nabla\cdot\boldsymbol{V}$ is the traceless rate of strain tensor and $\nu_\parallel$ and $\nu_{\text{iso}}$ are viscous diffusivities. Numerical particle diffusion can cause momentum and energy conservation errors. It is possible to add corrections to Eqs. (2-3) to preserve these conservation properties, as is done in NIMROD [45] and other codes.

NIMSTELL has options for solving either the magnetic field components or the magnetic vector potential in its time advance. The former solves the combined Faraday's/Ampere's equation with the resistive-MHD Ohm's law for electric field:

$$\frac{\partial}{\partial t}\boldsymbol{B}=\nabla\times\left(\boldsymbol{V}\times\boldsymbol{B}-\frac{\eta}{\mu_0}\nabla\times\boldsymbol{B}\right)+\kappa_b\nabla\nabla\cdot\boldsymbol{B}\,, \tag{6}$$

where $\eta$ is the electrical resistivity[1] and $\mu_0\boldsymbol{J} = \nabla\times\boldsymbol{B}$. The last term in (6) provides a diffusive correction to numerical divergence error [46], and it has proven effective in applications when used in combination with NIMROD's high-order spatial representation and flux-conserving boundary conditions [34]. NIMSTELL's alternative is to advance the vector potential using the resistive-MHD Ohm's law in the potential representation for electric field:

$$\frac{1}{\mu_0}\nabla\times\nabla\times\boldsymbol{A}-\underline{\boldsymbol{\sigma}}\cdot\left(\boldsymbol{V}\times\nabla\times\boldsymbol{A}-\frac{\partial}{\partial t}\boldsymbol{A}-\nabla\chi\right)=0\,. \tag{7}$$

Here, $\nabla\times\boldsymbol{A} = \boldsymbol{B}$, and $\underline{\boldsymbol{\sigma}}$ is the electrical conductivity tensor, which may have anisotropy with respect to $\hat{\boldsymbol{b}}$. The scalar field $\chi$ is a potential that is used to set a gauge condition. The analysis performed in Appendix A leads us to use $\nabla^2\chi = d_A\nabla\cdot\boldsymbol{A}$ to damp the divergence of the vector potential and approximately satisfy the Coulomb gauge.

Like NIMROD and other codes (see Ref. [47], for example), NIMSTELL's time advance is for nonlinear perturbations about a prescribed steady state without an amplitude ordering. This is not represented in Eqs. (1-7) for brevity. However, for example, the implemented continuity equation solves

$$\frac{\partial n}{\partial t}+\nabla\cdot(n_s\boldsymbol{V}+n\boldsymbol{V}_s+n\boldsymbol{V})=\nabla\cdot D_n\nabla n\,, \tag{8}$$

where $n_s$ and $\boldsymbol{V}_s$ are prescribed steady particle density and flow-velocity profiles that are assumed to satisfy time-independent equations, possibly including external sourcing. Applying NIMSTELL to linear time-dependent problems is achieved by not including the nonlinear terms, such as $n\boldsymbol{V}$ in Eq. (8). The separated steady fields are also convenient for nonlinear simulations of experimentally determined conditions, but one needs to consider physical consistency carefully if the prescribed fields do not satisfy the steady-state limit of solved evolution equations. Alternatively, a user may set the prescribed fields to zero or other steady states, such as vacuum magnetic field, and represent the rest in the evolved fields. NIMSTELL has preprocessing capability to load stellarator equilibria from the DESC code [48] into its steady-state fields with NIMSTELL's geometric mapping based on the shape of the DESC-computed flux surfaces. For slab and cylindrical test cases, steady-state fields and geometry are generated within NIMSTELL's preprocessor.

---

[1] Later, we also use "$\eta$" as an element coordinate, but meanings should be clear from context.

There is flexibility in the boundary conditions and initial conditions for NIMSTELL time-dependent computations. Particle density and temperature may be prescribed with Dirichlet conditions, or particle flux and conductive thermal flux may be imposed. Typically, flow-velocity is set to zero at the boundaries, though it is not a requirement of the model. When solving for vector potential with the *H*(curl) representation described in Section 3, our computations apply homogeneous Dirichlet conditions to the tangential components of $\boldsymbol{A}$ and to the scalar field $\chi$ to represent electrically conducting walls. However, our basis vectors for the magnetic representation do not generally align with 3D stellarator geometries, so we do not directly impose Dirichlet conditions on $\hat{\boldsymbol{n}} \cdot \boldsymbol{B}$. Instead, it is practical to include numerically implicit damping of $\hat{\boldsymbol{n}} \cdot \boldsymbol{B}$ in a weighted surface term—see Section 3.4.

## 3. Numerical methods

The primary distinction between time-dependent codes for stellarators and those for toroidally symmetric systems lies in their spatial representations. Nonetheless, we first briefly summarize NIMSTELL's temporal discretization.

### 3.1 Semi-implicit advance

The numerical methods used to advance NIMSTELL's solution fields over time are largely the same as those used in NIMROD. Flow-velocity is staggered one-half step in time relative to the particle density, temperature, and the magnetic field. MHD waves are stabilized at arbitrarily large timestep values ($\Delta t$) by the semi-implicit (SI) operator $\mathcal{L}_{\mathrm{si}}$ in the flow-velocity equation [49], [50]. It modifies Eq. (2) without loss of numerical consistency with the semi-discrete form of the advance of $\boldsymbol{V}$ from time level $j$ to $j+1$ being

$$\rho^{j+\frac{1}{2}}\left(1+\frac{\Delta t}{2}\boldsymbol{V}^j\cdot\nabla\right)\Delta\boldsymbol{V}+\frac{\Delta t}{2}\rho^{j+\frac{1}{2}}\Delta\boldsymbol{V}\cdot\nabla\boldsymbol{V}^j+f_\nu\Delta t\nabla\cdot\underline{\boldsymbol{\Pi}}(\Delta\boldsymbol{V})+C_{\mathrm{si}}\Delta t^2\mathcal{L}_{\mathrm{si}}(\Delta\boldsymbol{V})$$
$$=\Delta t\nabla\cdot\left[\frac{1}{\mu_0}\boldsymbol{B}\boldsymbol{B}-\underline{\mathbf{I}}\left(\frac{B^2}{2\mu_0}+p\right)\right]^{j+\frac{1}{2}}-\Delta t\rho^{j+\frac{1}{2}}\boldsymbol{V}^j\cdot\nabla\boldsymbol{V}^j-\Delta t\nabla\cdot\underline{\boldsymbol{\Pi}}\left(\boldsymbol{V}^j\right)\ . \tag{9}$$

Like NIMROD, NIMSTELL's semi-implicit operator is based on the linear ideal-MHD force operator, which draws from XTOR's formulation [51]. Applying the SI method to an underlying staggered leapfrog scheme, instead of XTOR's predictor-corrector, leads to wave propagation without dissipative numerical error [50], [44], so numerical smoothing must arise from dissipative terms in the system of equations. The anisotropies in NIMSTELL's SI operator are with respect to its 3D fields, which requires algebraic coupling among toroidal Fourier components in the advance of $\boldsymbol{V}$. The benefits of using a 3D SI operator has previously been considered for some NIMROD computations [38], but NIMROD computations typically use 2D SI operators to minimize Fourier coupling in the algebraic solves. In nonlinear computations, NIMSTELL's SI operator is updated over time to preserve its accuracy. Also, as indicated in Eq. (9), the viscous operator is implicit with numerical centering coefficient $f_v$, $0 \le f_v \le 1$, and advection is expanded one-half step from the previous time level [44].

### 3.2 Spatial representation

We consider 3D toroidal domains $\Omega$ by creating a non-overlapping partition of any poloidal plane into 2D finite elements and smoothly extending them over $\zeta$. The union of the partition includes the domain boundary $\partial\Omega$. Physical coordinates for each element have an isoparametric mapping by using the spatial expansion that is used for the expansions of $n$, $T$, and each component of $\boldsymbol{V}$, and for $\boldsymbol{B}$ when Eq. (6) is solved. With biquadratic and higher-order finite elements, element edges in a constant-$\zeta$ surface may be curved. Within each element $e$,

$$R_e(\xi,\eta,\zeta) = \sum_k \alpha_k(\xi,\eta)\left[R_{e_{k0}} + \sum_{n=1}^{N} R_{e_{kn}} e^{in\zeta} + cc.\right], \tag{10a}$$

$$Z_e(\xi,\eta,\zeta) = \sum_k \alpha_k(\xi,\eta)\left[Z_{e_{k0}} + \sum_{n=1}^{N} Z_{e_{kn}} e^{in\zeta} + cc.\right], \text{ and} \tag{10b}$$

$$\phi_e(\xi,\eta,\zeta) = \sum_k \alpha_k(\xi,\eta)\left[\phi_{e_{k0}} + \sum_{n=1}^{N} \phi_{e_{kn}} e^{in\zeta} + cc.\right] + \zeta \tag{10c}$$

with the element coordinates $0 \le \xi,\eta \le 1$ and $0 \le \zeta \le 2\pi$. The notation "$cc.$" refers to the complex conjugate of the previous term. The element bases $\alpha_k(\xi,\eta)$ for quadrilateral elements are the tensor products of the 1D Lagrange polynomials, $\lambda_i(\xi)$ and $\lambda_j(\eta)$, that have nodes at the points used for Gauss-Legendre-Lobatto (GLL) numerical integration. These polynomials form a space that is numerically equivalent to a finite expansion in Legendre polynomials while having the convenience of a nodal representation where coefficients represent function values at the node points [39], [40]. Ensuring function-value continuity along element interfaces just requires sharing coefficient values. For comparison, NIMROD's mapping does not have an expansion in $\zeta$. Apart from slab-geometry tests, we have not yet explored the potential benefits of distorting the physical angle $\phi$ over the generalized angle $\zeta$, so the $\phi_{e_{k0}}$ and $\phi_{e_{kn}}$ coefficients are 0 in the computations presented in Section 4. The expansion in Eq. (10) is also applicable to topologically toroidal but geometrically straight configurations through a different choice of Jacobian and metric computations.

The conductive heat-flux closure (4) makes the temperature equation (3) second-order with respect to spatial derivatives. This implies that the temperature field is in the *H*1 Hilbert space, where the function value and its first derivatives are square-integrable. The *H*1 space is also appropriate for the components of $\boldsymbol{V}$ when using the viscous closure (5) and for particle density when including the artificial particle diffusion in Eq. (1). When solving Eq. (6) with positive values of resistivity $\eta$ and divergence-error diffusivity $\kappa_b$, the magnetic field components are also in *H*1. Numerically, this representation allows nonzero magnetic divergence error, but with biquadratic and higher-order expansions, the diffusive correction term [46] is effective for magnetic-confinement applications [34]. It is also convenient to have all solution fields in the same space of functions.

Apart from the 3D geometry, the other significant difference between NIMSTELL and NIMROD is the option to solve the vector potential. This option uses the quadrilateral Nedelec edge elements [52] that are widely applied for computational electromagnetics to facilitate numerical convergence in the presence of sharp corners. Figure 1 shows a sketch of edge elements to help visualize the expansion. The union of these elements form a subspace of the ***H***(curl) space, where the vector functions and their curl are square-integrable. We combine 2D edge elements for the poloidal plane with 1D finite Fourier series over the generalized toroidal angle. In curvilinear-coordinate terms, the edge elements are expansions in reciprocal basis vectors of the mapping for a given element. Within each element, but without explicitly showing the Fourier expansion in $\zeta$,

$$\boldsymbol{A}_e(\xi,\eta,\zeta,t) = \sum_{k,l} A_{e,\xi_{k,l}}(\zeta,t)\varsigma_k(\xi)\lambda_l(\eta)\nabla\xi + \sum_{k',l'} A_{e,\eta_{k',l'}}(\zeta,t)\lambda_{k'}(\xi)\varsigma_{l'}(\eta)\nabla\eta$$
$$+ \sum_{k'',l''} A_{e,\zeta_{k'',l''}}(\zeta,t)\lambda_{k''}(\xi)\lambda_{l''}(\eta)\nabla\zeta \,. \qquad (11)$$

The $\lambda_k$ functions again represent 1D nodal polynomials based on GLL integration. However, as discussed in Appendix A, we consider expansions of different degree for $\boldsymbol{A}$ relative to those for the other physical fields. The $\varsigma_l$ functions are nodal polynomials based on Gauss-Legendre (GL) integration, where there are no nodes at the endpoints of the interval. The Nedelec edge elements have $\varsigma_l$ functions of one degree lower than those of the $\lambda_k$ functions. With the endpoint coefficients for the $\lambda_k$ bases of Eq. (11) being shared across element borders, tangential components of $\boldsymbol{A}(R,Z,\phi)$ are continuous regardless of the degree of the polynomial bases [53]. Another property of edge elements is that the normal component of their curl is continuous across element borders—see Appendix B. This satisfies the requirement of $\boldsymbol{B}\cdot\hat{\boldsymbol{n}}$ continuity across any interface. That $\nabla\cdot\boldsymbol{B}=0$ anywhere within an element follows directly from $\boldsymbol{B}=\nabla\times\boldsymbol{A}$ with $\boldsymbol{A}$ being defined everywhere within each $\boldsymbol{H}$(curl) element. Thus, this optional representation eliminates numerical $\nabla\cdot\boldsymbol{B}$ error to machine roundoff level.

When it is used, the NIMSTELL computation of $\boldsymbol{A}$ proceeds by solving Eq. (7) and the gauge relation in weak form. A step of the time advance determines the change in vector potential $\Delta\boldsymbol{A}\in\boldsymbol{A}_{h,N,p}$ and the scalar $\chi\in \mathrm{X}_{h,N,p}$ such that

$$\int dVol\left\{f_\eta \frac{\eta_s}{\mu_0}(\nabla\times\boldsymbol{G})\cdot\nabla\times\Delta\boldsymbol{A} + \boldsymbol{G}\cdot\left[\tilde{\sigma}_\delta\hat{\boldsymbol{b}}\hat{\boldsymbol{b}} + \tilde{\sigma}_\perp\underline{\boldsymbol{I}}\right]\cdot\left(\nabla\chi + \frac{1}{\Delta t}\Delta\boldsymbol{A}\right) - \frac{\tilde{\sigma}_\perp}{2}\boldsymbol{G}\cdot\boldsymbol{V}\times(\nabla\times\Delta\boldsymbol{A})\right.$$
$$\left. +\nabla\vartheta\cdot[\nabla\chi - f_d d_A\Delta\boldsymbol{A}]\right\}$$
$$= \int dVol\left\{\tilde{\sigma}_\perp\boldsymbol{G}\cdot\boldsymbol{V}\times(\nabla\times\boldsymbol{A}) - \frac{\eta_s}{\mu_0}(\nabla\times\boldsymbol{G})\cdot\nabla\times\boldsymbol{A} + d_A\nabla\vartheta\cdot\boldsymbol{A}\right\} \qquad (12)$$

for all $\boldsymbol{G}\in\boldsymbol{A}_{h,N,p}$ and all $\vartheta\in\mathrm{X}_{h,N,p}$. Here, $\boldsymbol{A}_{h,N,p}$ is the product subspace of 2D $\boldsymbol{H}$(curl) elements and the 1D Fourier exponential bases, and $\mathrm{X}_{h,N,p}$ is the continuous *H*1 subspace of 2D nodal spectral elements of the same degree as $\lambda_k$ and the 1D Fourier bases. The mesh spacing is represented by *h*, the polynomial degree is represented by *p*, and the largest toroidal Fourier wavenumber is *N*. We define a resistive scaling factor $\eta_s$ so that the electrical conductivities in Eq. (12) are normalized; $\sigma_\perp = \eta_s^{-1}\tilde{\sigma}_\perp$ and $\sigma_\parallel = \eta_s^{-1}\tilde{\sigma}_\parallel$. Also, $\tilde{\sigma}_\delta \equiv \tilde{\sigma}_\parallel - \tilde{\sigma}_\perp$. Although temporal indices are suppressed, the vector $\boldsymbol{A}$ is from level $j+1/2$, and with the staggering, $\boldsymbol{V}$ is at level $j+1$. The numerical coefficients $f_\eta$ and $f_d$ define the centering for resistive diffusion and for the gauge relation.

The low-resolution 2D mesh of elements shown in Fig. 2 is one that is used for the cylindrical interchange computation of Section 4.1. Topologically, it is typical of meshes used with NIMSTELL. The outer part is a polar mesh but near the axis, it transitions to a rectangular grid. We have not programmed triangular elements or degenerate quadrilateral elements, which would be needed to use a strictly polar mesh. The transitioning mesh also avoids extremely small element dimensions. When using DESC for stellarator equilibria, the outer topologically polar mesh is shaped to align with equilibrium magnetic flux surfaces. As also indicated by Fig. 2, the isoparametric mapping allows NIMSTELL to curve element sides.

### 3.3 Numerical interchange stabilization

When a numerical mesh is constructed and a set of basis functions is chosen, the spectral-element/Fourier representation for dependent variables yields a particular subspace of the *H*1 space. The time-dependent computation then determines, by some means, the best set of functions from a given subspace at each step. However, it is well known that difficulties arise in hyperbolic problems when attempting to model configurations with insufficient numerical resolution for a given set of dissipation coefficients, see Refs. [54], [55], and [39] for example. In these conditions, finite-element "stabilization" methods [56], [57] have proven effective for many applications, including MHD [58], [59], [60], [61].

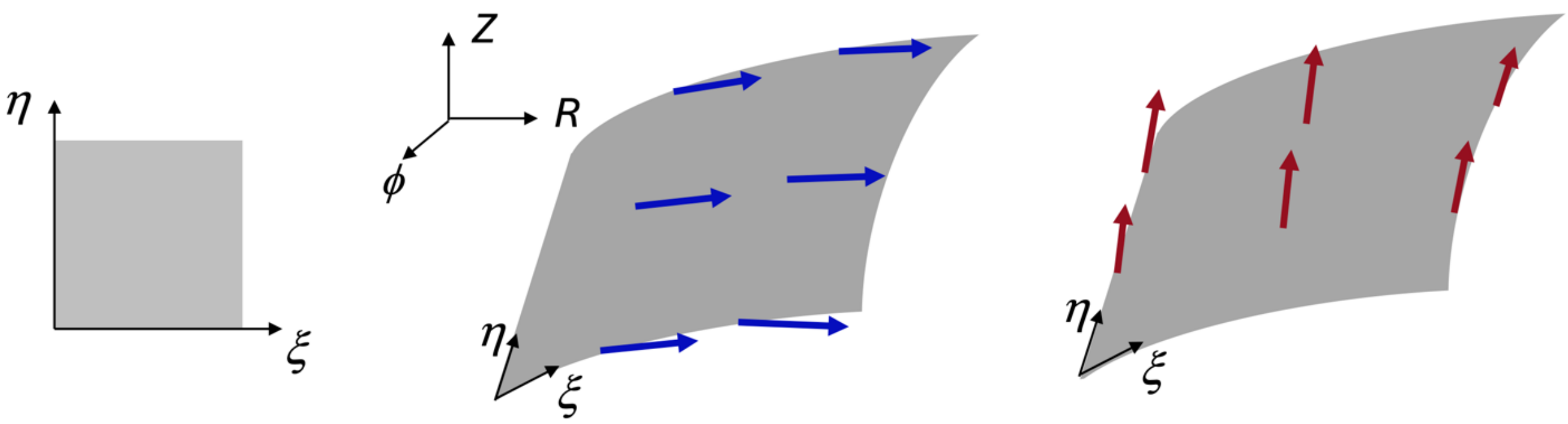

Figure 1. Sketches of the reference element with $(\xi, \eta)$ element coordinates (left) and the $\nabla\xi$ (center) and the $\nabla\eta$ (right) basis vector functions after mapping to physical coordinates.

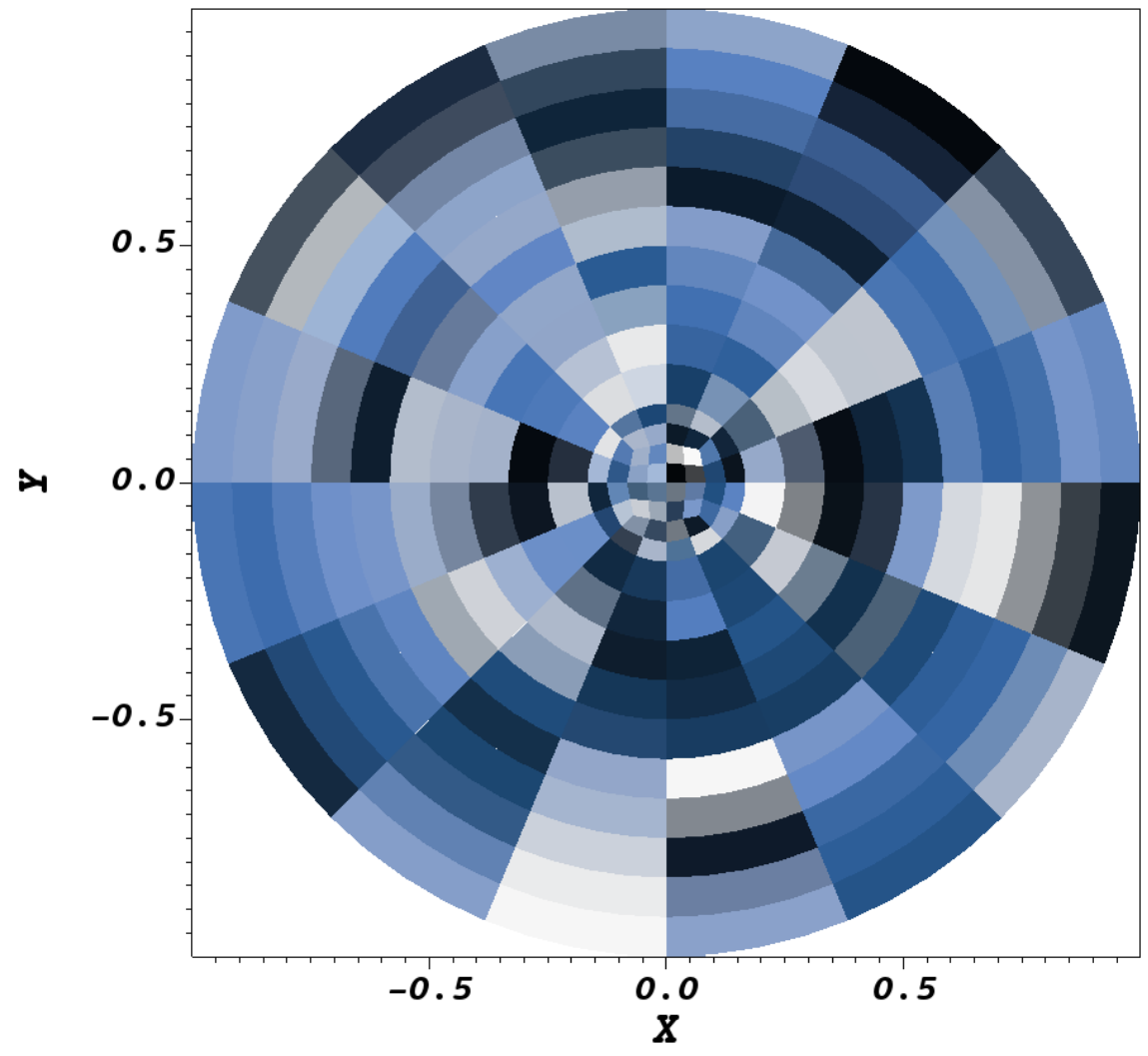

Figure 2. Finite-element mesh of the lowest-resolution cylindrical interchange computation. The elements are curved and transition from a polar grid (12 radial × 16 azimuthal) to a rectangular grid of 16 elements at the axis.

At large $\Delta t$-values, the second-order derivatives on $\Delta \boldsymbol{V}$ in NIMROD's semi-implicit operator enhance the coercivity of the computation for $\Delta \boldsymbol{V}$, itself. However, without further stabilization, the time-advance converges on resonant interchange from the unstable side, meaning that numerically computed growth rates are larger at low spatial resolution than their converged values. As noted in Ref. [62], this is problematic for nonlinear time-dependent computations when physical modes saturate but growing numerical modes arise. For NIMROD, we previously developed specialized stabilization terms to make the representation converge from the stable side when using equal-order expansions of all physical fields [63].

The parabolic form of these terms is implemented in NIMSTELL for use with its equal-order $H1$ representation of magnetic-field components. At each step the flow-velocity advance finds $\Delta\boldsymbol{V} \in \boldsymbol{V}_{h,N,p}$ band $\sigma, \varpi \in F_{h,N,p}$ such that

$$\begin{aligned}
&\int dVol\left\{\rho\boldsymbol{w}^* \cdot \Delta\boldsymbol{V} + \Delta t\boldsymbol{w}^* \cdot \left(\frac{\rho}{2}\boldsymbol{V}\cdot\nabla\Delta\boldsymbol{V} + \frac{\rho}{2}\Delta\boldsymbol{V}\cdot\nabla\boldsymbol{V}\right) + \Delta t(\upsilon^*\sigma + \mu^*\varpi)\right.\\
&\quad + C_{\mathrm{si}}\Delta t^2\left[\frac{1}{\mu_0}\nabla\times(\boldsymbol{w}^*\times\boldsymbol{B})\cdot\nabla\times(\Delta\boldsymbol{V}\times\boldsymbol{B}) + \Gamma p(\nabla\cdot\boldsymbol{w}^*)(\nabla\cdot\Delta\boldsymbol{V})\right]\\
&\quad - \frac{C_{\mathrm{si}}\Delta t^2}{2}\left[\boldsymbol{w}^*\cdot\boldsymbol{J}\times\nabla\times(\Delta\boldsymbol{V}\times\boldsymbol{B}) + \Delta\boldsymbol{V}\cdot\boldsymbol{J}\times\nabla\times(\boldsymbol{w}^*\times\boldsymbol{B}) - \left((\nabla\cdot\boldsymbol{w}^*)\Delta\boldsymbol{V} + (\nabla\cdot\Delta\boldsymbol{V})\boldsymbol{w}^*\right)\cdot\nabla p\right]\\
&\quad \left. - f_v\Delta t\nabla\boldsymbol{w}^H : \underline{\boldsymbol{\Pi}}(\Delta\boldsymbol{V})\right\}\\
&= \int dVol\left\{\Delta t\left[\boldsymbol{w}^*\cdot\boldsymbol{J}\times\boldsymbol{B} + p\nabla\cdot\boldsymbol{w}^* + \nabla\boldsymbol{w}^H : \underline{\boldsymbol{\Pi}}(\boldsymbol{V}) - \rho\boldsymbol{V}\cdot\nabla\boldsymbol{V}\cdot\boldsymbol{w}^*\right]\right.\\
&\quad \left. + \sqrt{d_d P_c \Delta t}(\upsilon^*\nabla\cdot\boldsymbol{V} - \sigma\nabla\cdot\boldsymbol{w}^*) + \sqrt{d_v\Delta t}\boldsymbol{B}\cdot(\mu^*\nabla\times\boldsymbol{V} - \varpi\nabla\times\boldsymbol{w}^*)\right\} \qquad (13)
\end{aligned}$$

for all $\boldsymbol{w} \in \boldsymbol{V}_{h,N,p}$ and $\upsilon, \mu \in F_{h,N,p}$. Here, $\boldsymbol{V}_{h,N,p}$ is the continuous $H1$ subspace. We also define $P_c \equiv B^2 + \Gamma p$ as the total pressure for compressional waves. Setting homogeneous essential conditions on $\boldsymbol{V}$ along the domain surface eliminates surface terms that would result from integration by parts. As described in [63], $F_{h,N,p}$ is an incomplete space of discontinuous polynomials in the meshed plane with just the $p$-th Legendre polynomial along at least one element side. Including the $\sigma\nabla\cdot\boldsymbol{w}^*$ and $\varpi\nabla\times\boldsymbol{w}^*$ projections in the $\Delta\boldsymbol{V}$-equation does not affect numerical consistency, since the projections onto the incomplete $F_{h,N,p}$ space vanish with increasing resolution of sufficiently smooth functions. The dimensionless coefficients $d_d$ and $d_v$ are of order 1. In the limit of small viscosity, the $\sigma$-projection provides a response to $\nabla\cdot\boldsymbol{V}$ at the scale of the nodes, and the $\varpi$-projection provides a response to parallel vorticity at that scale, which yields the desirable converge from the stable side. In comparison with Eq. (12) of Ref. [34] and Eq. (20) of Ref. [63], NIMSTELL uses complete 3D fields in the interchange-stabilizing projections and in its semi-implicit operator.

Comparing the flow-velocity equation for the two magnetic-field options, the contribution of the Lorentz force density on the right side of (13) is expressed as implemented in NIMROD and in NIMSTELL for the $H1$ representation of magnetic field components. With components of $\boldsymbol{B}$ in $H1$, current density may be evaluated from $\nabla\times\boldsymbol{B}$ at the numerical integration points. For the $\boldsymbol{H}$(curl) representation of vector potential, we use the divergence of the Maxwell stress tensor, as shown in Eqs. (2) and (9). This divergence is integrated by parts when projected onto the test vector $\boldsymbol{w}$, since differentiation of $\boldsymbol{A}$ is limited to a single curl operation. However, $\boldsymbol{J}$ and $\nabla\boldsymbol{B}$ are needed to update the semi-implicit operator in nonlinear computations. Thus, when using $\boldsymbol{A}$ in $\boldsymbol{H}$(curl) in nonlinear computations, an extra projection of $\boldsymbol{B}$ components onto an $H1$ expansion is performed during each time advance for this purpose. The interchange-stabilization terms are not used when selecting the vector-potential option.

### 3.4 Magnetic-field surface term

The $H1$ representation of magnetic-field components uses cylindrical unit vectors $(\hat{\boldsymbol{R}}, \hat{\boldsymbol{Z}}, \hat{\phi})$ with respect to the geometric axis through the center of a torus. For stellarator configurations, the outward surface normal vector $\hat{\boldsymbol{n}}$ generally has three nonzero cylindrical components. Taking $\xi$ to be the element coordinate that is tangent to the domain boundary in an element that is adjacent to the boundary, $\hat{\boldsymbol{n}} \propto \partial\boldsymbol{R}/\partial\xi \times \partial\boldsymbol{R}/\partial\zeta$. In NIMSTELL, the mapping has a Fourier expansion over $\zeta$, so it is not convenient to impose $\hat{\boldsymbol{n}} \cdot \boldsymbol{B} = 0$ along the surface through essential conditions on components of $\boldsymbol{B}$ without setting all components to 0, which is too restrictive. To approximate the $\hat{\boldsymbol{n}} \cdot \boldsymbol{B} = 0$ Dirichlet condition, we instead add a surface term that rapidly damps $\hat{\boldsymbol{n}} \cdot \boldsymbol{B}$ along the surface. The formulation for advancing $\boldsymbol{B}$ is to find $\Delta\boldsymbol{B} \in \boldsymbol{V}_{h,N,p}$ such that

$$\int dVol \left\{ \boldsymbol{C}^* \cdot \Delta\boldsymbol{B} + \Delta t \nabla \times \boldsymbol{C}^* \cdot \left( f_\eta \frac{\eta}{\mu_0} \nabla \times \Delta\boldsymbol{B} - \frac{1}{2} \boldsymbol{V} \times \Delta\boldsymbol{B} \right) + f_{\kappa_b} \Delta t \kappa_b (\nabla \cdot \boldsymbol{C}^*)(\nabla \cdot \Delta\boldsymbol{B}) \right\}$$
$$+ \oint dS[(1 + d_S)(\hat{\boldsymbol{n}} \cdot \boldsymbol{C}^*)(\hat{\boldsymbol{n}} \cdot \Delta\boldsymbol{B})]$$
$$= \Delta t \int dVol \left\{ \nabla \times \boldsymbol{C}^* \cdot \left( \boldsymbol{V} \times \boldsymbol{B} - \frac{\eta}{\mu_0} \nabla \times \boldsymbol{B} \right) - \Delta t \kappa_b (\nabla \cdot \boldsymbol{C}^*)(\nabla \cdot \boldsymbol{B}) \right\} - \oint dS[d_S(\hat{\boldsymbol{n}} \cdot \boldsymbol{C}^*)(\hat{\boldsymbol{n}} \cdot \boldsymbol{B})] \quad (14)$$

for all $\boldsymbol{C} \in \boldsymbol{V}_{h,N,p}$, the same subspace used for components of $\boldsymbol{V}$. For $d_s > 1$ but not so large as to make the resulting matrix ill-conditioned, the surface term penalizes $\hat{\boldsymbol{n}} \cdot \boldsymbol{B}$ along the surface. In practice, many of our computations set $d_s \sim 10$. We remark that $\hat{\boldsymbol{n}}$ in toroidally symmetric computations also generally does not align with either $\hat{\boldsymbol{R}}$ or $\hat{\boldsymbol{Z}}$. Setting $\hat{\boldsymbol{n}} \cdot \boldsymbol{B} = 0$ along the domain surface of NIMROD computations is accomplished through local constraint relations at each finite-element node along the surface. The distinction is that with $\hat{\boldsymbol{n}} \cdot \hat{\boldsymbol{R}}$ and $\hat{\boldsymbol{n}} \cdot \hat{\boldsymbol{Z}}$ independent of $\phi$, the NIMROD constraints are imposed separately for each toroidal Fourier component of $\boldsymbol{B}$.

We also contrast this surface term for the $H1$ representation of magnetic components with the essential condition imposed on the tangential components of $\boldsymbol{A}$ for conducting surfaces when Eq. (12) is used. Staying with the convention of the logical coordinate $\eta$ being constant along the domain surface for adjacent elements, $\partial\boldsymbol{R}/\partial\xi$ and $\partial\boldsymbol{R}/\partial\zeta$ are tangent to the surface. With the $\boldsymbol{H}$(curl) bases, setting $\hat{\boldsymbol{n}} \times \boldsymbol{A} = 0$, is accomplished by setting the expansion coefficients $A_{e,\xi_{k,l}}$ and $A_{e,\zeta_{k,l}}$ in Eq. (11) to 0 for nodes along the surface, for all Fourier components, and for all time. Setting the coefficients of $\chi$ to zero, or any unique value, along the surface then ensures $\hat{\boldsymbol{n}} \times \boldsymbol{E} = 0$.

### 3.5 Algebraic solves

NIMSTELL uses preconditioned conjugate gradients (CG) and generalized minimal residual (GMRES) Krylov-space methods to solve the linear algebraic systems for the implicit and semi-implicit time-advance equations. With 3D geometry, the algebraic systems have coupling among the Fourier components, even for linearized systems of equations. Because the complete matrix is too large to form in general, NIMSTELL computes the implicit operations on the Krylov-space vectors through the "matrix-free" method. This method rearranges the order of the summations for numerical integration and matrix-vector products. The associated computational cost is performing finite-element operations at each step of the Krylov algorithm. Preconditioning applies approximate inverses at each step to accelerate convergence and is critical for making Krylov methods practical for large ill-conditioned systems [64]. The NIMSTELL implementation of CG and GMRES is similar to NIMROD's, but the computational infrastructure to use the more general data layout for coefficients of $\boldsymbol{H}$(curl) expansions has been added.

Preconditioning operations in NIMROD are based on inverting non-overlapping submatrices ("blocks") that cover different levels of coupling over the poloidal-plane but are diagonal in the toroidal Fourier basis. Typically, the most effective approach is to solve the sparse matrices that represent coupling over the entire (poloidal) plane of elements using the parallel SuperLU_DIST library solver [65]. The strong coupling among Fourier components from 3D stellarator geometry makes the block Fourier diagonal (bFd) approach ineffective for general NIMSTELL application. To address this, the bFd method has been expanded to include multiple toroidal Fourier components in the blocks that are solved by the direct solver. The programming allows the blocks to either cover bands of Fourier components or user-defined families of Fourier components. The implementation allows overlap of the blocks, meaning that a Fourier component can be included in more than one block. The preconditioner averages the results of the block solves for the Fourier components that are included in more than one block for additive-Schwarz preconditioning [64].

Using preconditioner blocks that span continuous ranges of toroidal Fourier components is effective for single field-period ($N_p = 1$) stellarator configurations, but the effectiveness decreases with increasing $N_p$ values. This can be appreciated from the illustrations shown in Fig. 3. The information on matrix-element scaling represents an $N_p = 5$ stellarator, and Fig. 3a shows that using preconditioning blocks for continuous ranges wastes computational effort and memory on negligible couplings. However, using blocks that include just the strongest couplings that result from the symmetry of the configuration (Fig. 3b) is effective. Table I compares the performance of these block preconditioning choices on a tearing-mode computation for an $N_p = 5$ configuration with elliptical cross section, the W7-A case of Ref. [32] and Section 4.2. For this linear computation, the judicious blocks of $n$-values that are coupled by the symmetry of the geometry and the equilibrium is as effective as using one matrix covering all Fourier components while requiring seven times less memory and allowing faster computation with the smaller matrices.

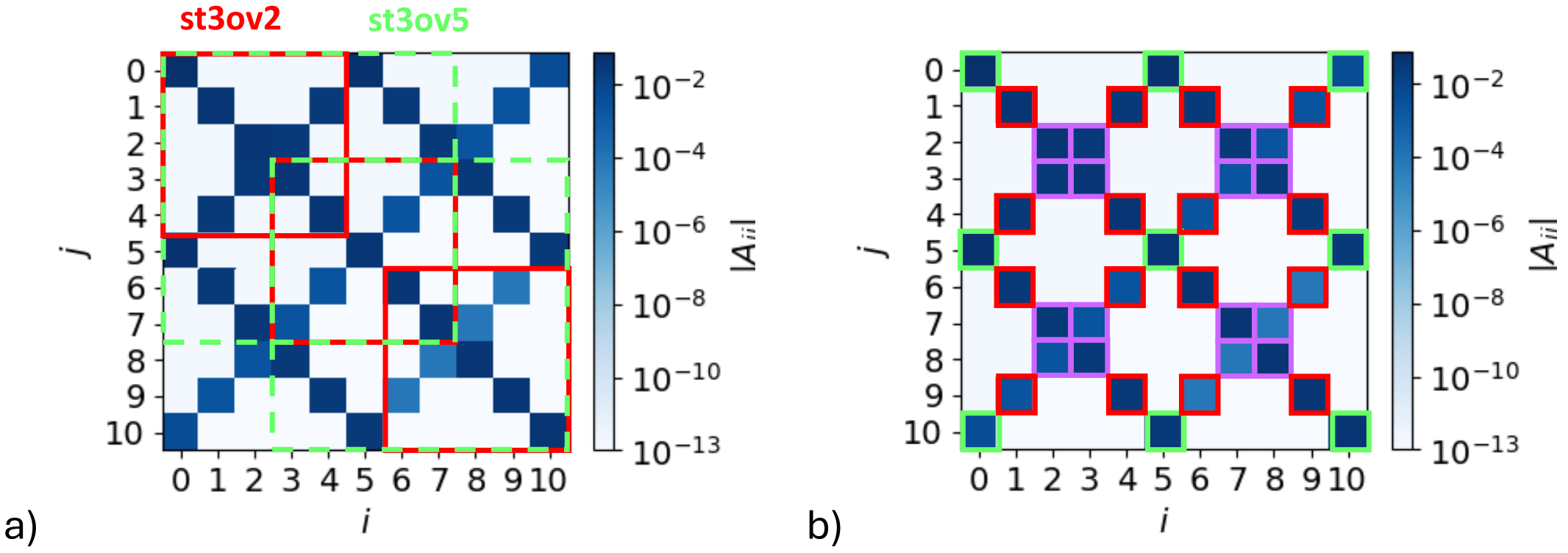

Figure 3. Illustration of the magnitude of matrix elements for the $N_p = 5$ case, toroidal resolution of $0 \le n \le 10$, and different preconditioner block choices. The allocation of preconditioning blocks in a) represents two stride=3 (“st3”) sets of blocks with overlap (“ov”) of 2 (red) or 5 (green). The allocation of preconditioning blocks in b) is user-prescribed according to the families (red, green, and violet) of $n$-values that have either $n_1 + n_2$ or $n_1 - n_2$ being a multiple of $N_p$, where $n_1$ and $n_2$ are two of the same family.

Table I. Comparison of Krylov-space solver performance on the $N_p = 5$ tearing-mode computation when preconditioning with the different Fourier block decompositions shown schematically in Fig. 3.

| Block Composition ($n$) | Iteration Count | Memory Use |
|---|---|---|
| {0-4}, {3-7}, {6-10} | ~15,000 | 550 GB |
| {0-7}, {3-10} | ~2,000 | 1.4 TB |
| {0-10} | 2 | 2.4 TB |
| {1,4,6,9}, {2,3,7,8}, {0,5,10} | 2 | 350 GB |

**4. Test applications**

The applications presented in this article test resonant interchange in cylindrical geometry, magnetic tearing in the toroidal W7-A geometry, and magnetic tearing in a quasi-axisymmetric configuration. We also demonstrate convergence on anisotropic thermal conduction in a straight stellarator configuration.

4.1 Linear resonant interchange

Resonance for low-frequency dynamics is the condition that helical wavefronts align with the equilibrium magnetic field so that the wavenumber vector, $\boldsymbol{k} = mr^{-1}\hat{\boldsymbol{\theta}} + 2\pi n L_z^{-1}\hat{\boldsymbol{z}}$ in cylindrical geometry, satisfies $\boldsymbol{k} \cdot \boldsymbol{B} = 0$. To verify our implementation of the numerical stabilization terms in Eq. (13) and to demonstrate performance of the vector-potential option, we consider the same cylindrical interchange instability from Ref. [2] that was used in Ref. [63]. The parameterized 1D equilibrium profile has uniform axial field and mass density with $B_z = 1$ and $\rho = 1$ after normalization. Axial current density leads to the cylindrical safety-factor profile of

$$q(r) = \frac{2\pi r B_z(r)}{L_z B_\theta(r)} = 2\pi \frac{1+c_2^2 r^2}{L_z c_1}$$

over $0 \le r \le 1$, where $q$ is the number of axial passes per azimuthal pass of a field-line and $L_z$ is the length of the periodic cylinder. The Suydam parameter for resonant ideal interchange stability in these profiles is

$$D_s(r) \equiv -\frac{2\mu_0}{r B_z^2}\left(\frac{q}{dq/dr}\right)^2 \frac{dP}{dr} = \frac{c_1^2}{(1+c_2^2 r^2) c_2^4 r^2} \ ,$$

and the criterion for stability is $D_s \le 1/4$. For parameters with $c_1 = 4/7$ and $c_2 = 10/7$, the helical mode with azimuthal wavenumber $m = 4$ and axial wavenumber $n = -1$ for $L_z = 3.523$ is resonant along the cylindrical surface defined by $r = 0.371$ where $D_s = 0.443$. Our reference value of the growth rate is $\gamma_{\text{eig}} = 6.771 \times 10^{-3} \tau_a^{-1}$, where $\tau_a = a\sqrt{\mu_0 \rho}/B_z$. This value is computed from a 1D ideal-MHD eigenvalue solver [63] applied to magnetically aligned components of the displacement vector [2] for the selected helical mode. We have used Richardson extrapolation to refine the eigenvalues $(-\gamma_{\text{eig}}^2)$ computed with 1280 and 2560 linear elements over $r$.

All physical dissipation parameters and $D_n$ are effectively 0 in the NIMSTELL computations considered here. To avoid artificial magnetic reconnection, we also use a relatively small normalized value of $\kappa_b = 10^{-5}$. With the chosen $\Delta t$ values of 5 for the mesh-spacing scan and 10 for the polynomial-degree scan, diffusion for divergence correction per step is on the spatial scale of $10^{-2}$. As a check, reducing $\kappa_b$ to $10^{-6}$ in biquartic computations with 128 azimuthal elements over the azimuthal angle decreases the growth rate by 0.2%. Although this form of error control has greater effect at coarser resolution (for example, 0.7% at 64 azimuthal elements), it does not adversely affect our convergence tests. With respect to temporal resolution, spatially resolved results for $\Delta t = 5$ and 10 agree to within 0.015%; at $\Delta t = 40$ the error is only 0.4% for this ideal mode. For the stabilization terms in Eq. (13), we use the standard NIMROD coefficients of $d_d = 1$ and $d_v = 0.1$.

Our results show that NIMSTELL converges from the stable side when refining either with respect to mesh spacing or with respect to the polynomial degree of the spectral-element basis functions. The error shown in Fig. 4 is $(\gamma_{\text{eig}} - \gamma_{\text{NIM}})\tau_a$, and all errors are positive, hence the NIMSTELL growth rates are smaller than the converged value. The meshes have the same distribution of elements shown in Fig. 2. The horizontal axis of Fig. 4a indicates the number of elements in the azimuthal direction of the polar region with the number of elements in the radial direction varying proportionately. A biquartic computation with the 208-element mesh of Fig. 2 (16 over the azimuthal angle) does not produce a growing mode over $10^4\ \tau_a$. A mesh with four times the number of elements is the lowest-resolution computation of Figs. 4a-b. Figure 4a also shows three results computed with NIMROD. NIMSTELL yields the same growth rates to at least three significant digits, confirming NIMSTELL's implementation of the interchange-stabilization terms. We comment that even for an axisymmetric mesh and equilibrium, the implementations differ. The NIMSTELL computations minimally need Fourier components $0 \le n \le 1$ and use Fourier transforms for numerical integrals on the $\zeta$-grid, whereas NIMROD computes linear results for $n = 1$ alone without Fourier transforms. Another NIMSTELL computation shown in Fig. 4a has its mesh twisting a full turn over $\zeta$. In this case, the equilibrium and the mesh have $n > 0$ components. The eigenmode shape only twists 1/4 of a turn over $2\pi$ in $\zeta$, so the mesh and the mode are intentionally misaligned. The solution then has Cartesian $x$ and $y$ vector components in the $n = 2$ and $n = 4$ Fourier components, and the scalar fields and axial vector components are in $n = 3$. Its growth rate matches the NIMSTELL value with the axisymmetric mesh.

The eigenfunction for this mode is characterized by its sharp features at the resonance (see Fig. 15b of Appendix A) . Conditions are close to marginal stability, and only inertia keeps components of flow-velocity from being discontinuous. Thus, the convergence properties of the representation are not optimal for the eigenmodes. For sufficiently smooth functions, $C^0$ biquartic elements will achieve fifth-order convergence—see Theorem 3.3 of Ref. [54]. The lower rate observed from Fig. 4a can be expected from the nearly singular behavior of the eigenfunction. The roughly linear behavior of error in the polynomial-degree scan plotted on the semi-log scale of Fig. 4b is suggestive of geometric convergence [40], but one should be skeptical with the limited range of results. In fact, it is surprising that this spectral-resolution scan yields the accuracy that it does, given the application. We also emphasize that although the location of the resonant surface is known *a priori*, no mesh packing is used to represent application of NIMSTELL without such prior knowledge.

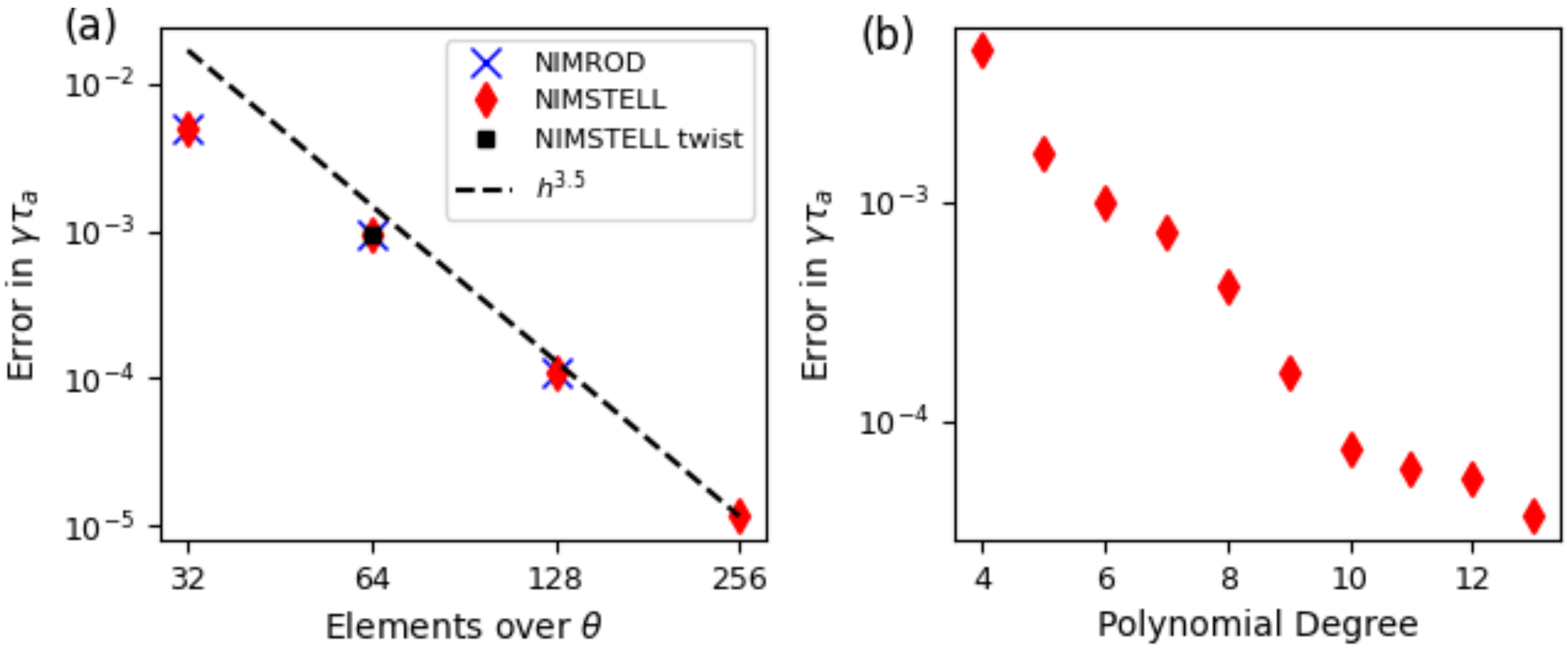

Figure 4. Convergence of NIMSTELL on the $m = 4, n = 1$ cylindrical interchange for $L_z = 3.523$ using the $H$1 magnetic-component option. Convergence is shown with respect to (a) mesh spacing for biquartic elements and (b) polynomial degree of the spectral-element basis functions for a mesh of 832 elements (32 over $\theta$). Results from NIMROD and a single NIMSTELL computation with a twisting mesh are also plotted in (a), together with a power-law scaling of mesh spacing.

We find that the same set of ideal-MHD computations are not tractable with NIMSTELL's present vector-potential implementation. With $\eta = 0$ in this ideally unstable case, numerical noise grows at $\gamma\tau_a \approx 0.03$ with biquadratic elements and meshes of either 32 or 64 elements over $\theta$. A minimum amount of electrical resistivity is apparently needed for a given mesh. To show this, we consider the resistive version of the ideally unstable conditions and resistive interchange in conditions that are ideally stable. The latter has $c_1 = 1/2$, $c_2 = 5/4$, and $L_z = 2\pi$ to make the $(3,1)$ mode resonant at $r = 0.566$, where $D_s = 0.213$; noise also grows in an $\eta = 0$ computation with this equilibrium. In these computations, we vary $\eta$ starting with a relatively large value of $10^{-4}$ and a coarse mesh and decrease $\eta$ until the growth rate is inaccurate or produces a noisy result. The mesh is then refined, and the resistivity scan is continued.

As shown in Fig. 5a for the ideally unstable case, computations using biquartic elements with quintic $\lambda_k$ for the vector- and scalar-potential expansion (11) and 64 elements over the azimuthal direction produce reliable results for $\eta \geq 10^{-7}$. At $\eta = 10^{-8}$, the numerical eigenmode has noise. Computations with 128 elements over $\theta$ are reliable through $\eta = 10^{-8}$, but produce rapidly growing noise at $\eta = 10^{-9}$. A computation with 256 elements over $\theta$ at $\eta = 10^{-9}$ produces the expected localized eigenmode with $\gamma$ being 1.2% larger than the reference ideal value. In these ideally unstable conditions, resistive reconnection at the resonant surface allows faster than ideal growth rates at large-$\eta$ [66]. In the ideally stable conditions (Fig. 5b), the resolved computations follow the scaling of $\gamma \sim \eta^{1/3}$ that is expected from asymptotic theory for resistive interchange [67]. However, we again find cases where the resistive diffusion is insufficient to avoid noise at a given level of spatial resolution.

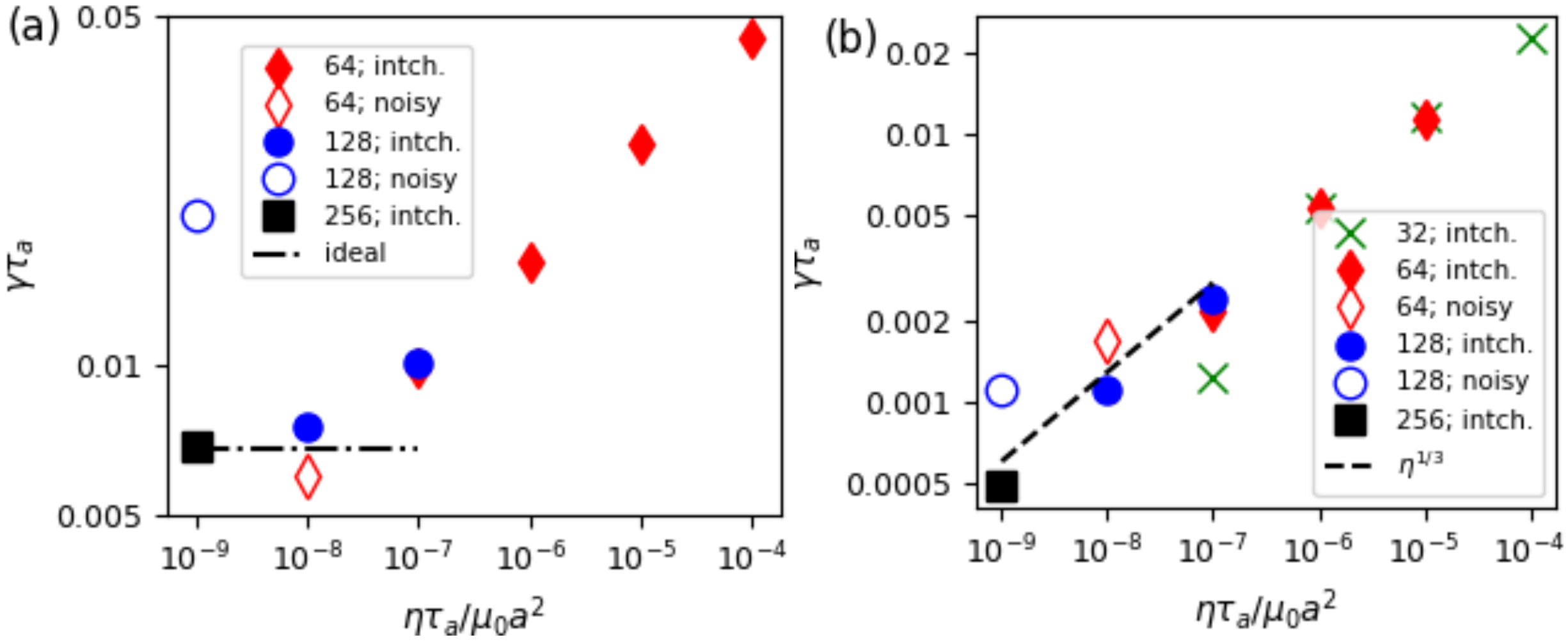

Figure 5. Resistive scaling of NIMSTELL on the $m = 4, n = 1$ cylindrical interchange using the ***H***(curl) vector-potential option. Results are for (a) the ideally unstable $(m = 4, n = 1)$ $L_z = 3.523$ case and (b) the ideally stable $(m = 3, n = 1)$ $L_z = 2\pi$ case. The two series of computations have the same basis functions but different meshes, as indicated by the number of elements over $\theta$.

Where the interchange computations with vector potential do not produce noise, under-resolved computations grow more slowly than converged results. Momentarily ignoring the noisy results, this option also converges from the stable side, which is consistent with the numerical eigenvalue analysis with 1D meshing discussed in Appendix A. We infer that the 1D analysis is not capable of identifying the issue that leads to growing noise. A plausible explanation is simply that the resistive diffusion term, the first term on the left of Eq. (12), is needed to make the ***H***(curl) vector-potential formulation coercive. Appendix A describes numerical eigenmodes that do not couple to flow or excite other response for ideal MHD with this representation, and their existence is related. While the damping added for the gauge condition contributes to coercivity and removes other numerical modes in the ideal-MHD analysis, it is not sufficient for reliable numerical algebra when resistive diffusion is minimized.

To check nonlinear evolution with the *H*1 magnetic-field representation, we compare NIMSTELL and NIMROD results on the profile with $c_1 = 4/7$ and $c_2 = 10/7$, but $L_z = 4\pi/3$, which has its marginal ideal stability point at $r = 0.466$. The $(3,1)$ mode is resonant at $r = 0.264$ and is the only $n = 1$ mode that is ideally unstable. For $n = 2$, the $(6,2)$ and $(7,2)$ modes are ideally unstable. Although larger-$n$ wavenumbers have more unstable modes, the smaller wavelengths lead to smaller growth rates with viscous dissipation in the system, and they are driven primarily by nonlinear coupling that takes energy from the dominant modes. Both computations use the numerical stabilization method described in Section 3.3. We set $\mathrm{S} = \tau_r/\tau_a = \mu_0 a^2/\eta\tau_a = 10^8$ and $\mathrm{Pm} = \mu_0\nu_{\mathrm{iso}}/\eta = 10$. There is also a small level of thermal conduction with $\kappa_{\mathrm{iso}}$ set to $nk_B(\Gamma - 1)\nu_{\mathrm{iso}}$. The particle density is initially uniform, so the pressure gradient is in the temperature profile. For numerical purposes, we set $D_n = \nu_{\mathrm{iso}}$, and the normalized $\kappa_b = 0.01$. The two computations use the same mesh of $31 \times 40$ elements over an annular region surrounding 100 at the axis. The basis functions are biquintic, and the Fourier expansion over $z$ retains $0 \le n \le 21$. Also, although it is not a steady state for all equations, we retain the MHD equilibrium in the steady-state fields of our expansion, as if unspecified sources maintain it. Symmetric $(n = 0)$ distortions develop in the computations through nonlinear effects, not through dissipation.

Figure 6a-b compares the NIMROD and NIMSTELL magnetic and kinetic fluctuation energies for $0 \le n \le 4$. The traces lie on top of each other through the initial nonlinear saturation phase and begin to deviate after that. Larger-$n$ harmonics have a similar level of agreement. Figure 6c shows the total pressure in the $z = 0$ plane at $t = 500\ \tau_a$. After this time, the evolution becomes turbulent. Quantitative discrepancies then accumulate, possibly due to the iterative solver tolerance, despite the fact that the two codes apply the same numerical methods.

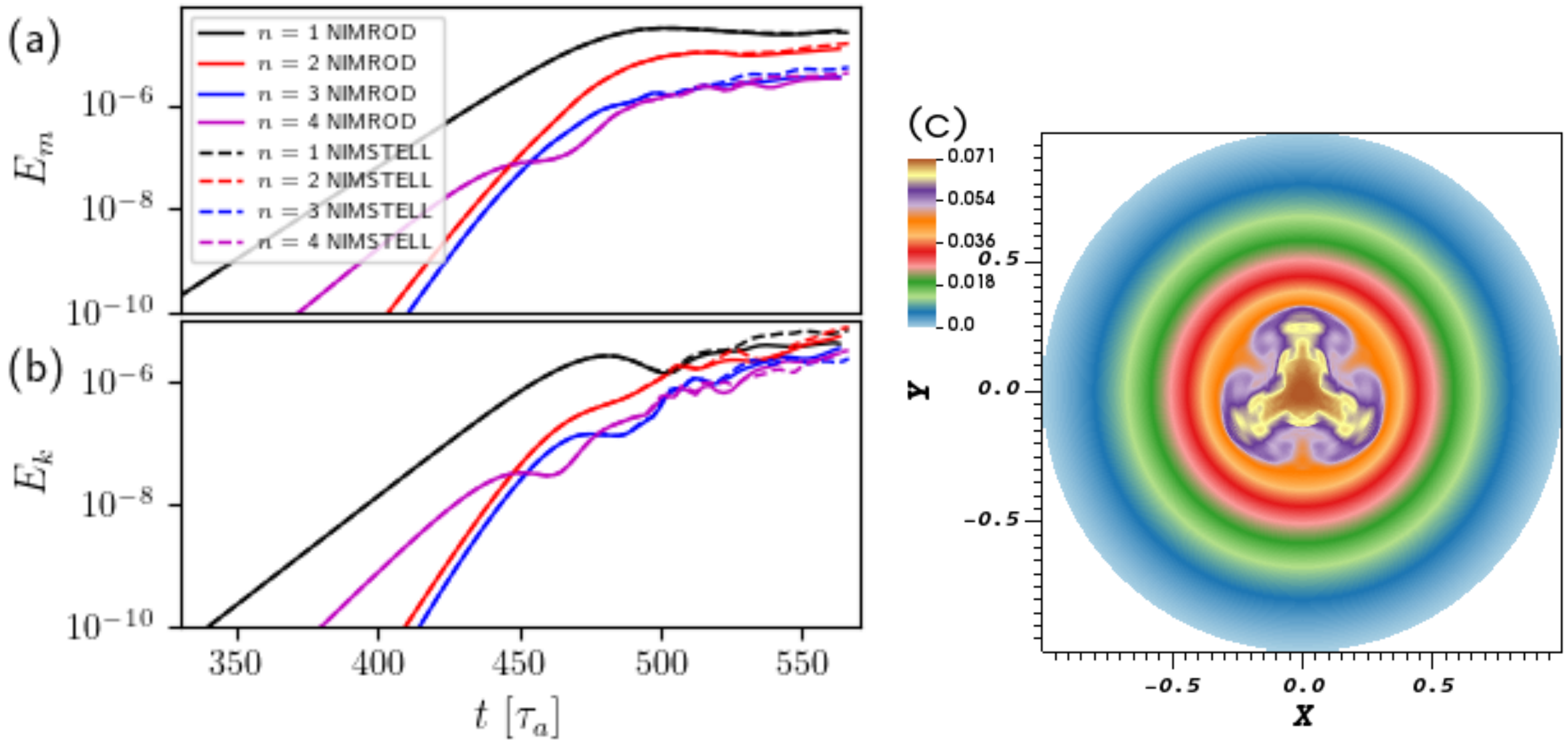

Figure 6. Comparison of fluctuation energies (a-b) from NIMROD and NIMSTELL computations for $1 \le n \le 4$ of the nonlinear interchange computation and (c) the sum of perturbed and equilibrium pressure in the $z = 0$ plane from the NIMSTELL computation at $t = 500\ \tau_a$. For reference, the energy in the equilibrium magnetic field is 6.8 in this normalized configuration.

### 4.2 Resistive tearing

We have verified NIMSTELL on linear resistive tearing in cylindrical geometry, but here we focus on toroidal stellarator computations. The first configuration represents the W7-A stellarator [68], where the elliptical cross section rotates over the toroidal angle with the toroidal field-period $(N_p)$ of five. The aspect ratio is large with the toroidal major radius $R = 1.99$ m and major and minor axes of the ellipse being 0.1178 m and 0.09131 m. The experiment had an Ohmic transformer, and toroidal current was induced in the plasma in some discharges. We use the profile specifications from Ref. [32], as applied for the JOREK code, repeated here for convenience:

$$I_n(\psi_{tn}) = 3\psi_{tn} - 3\psi_{tn}^2 + \psi_{tn}^3 \text{ and}$$

$$P(\psi_{tn}) = P_a - (P_a - P_b)\psi_{tn}\ ,$$

where $\psi_{tn}$ is the normalized toroidal magnetic flux function that varies from 0 on the magnetic axis to 1 at the plasma surface, $I_n$ is the toroidal current normalized by 17.5 kA, and for this low plasma-$\beta$ case $P_a = 1$ Pa and $P_b = 0.01$ Pa. Here, we consider just one set of dissipation coefficients: $\eta = 1.938 \times 10^{-6}\ \Omega\text{m}$, $\text{Pm} = 0.001$, $\nu_\parallel/\nu_{\text{iso}} = 1000$, and $(\Gamma - 1)\kappa_{\text{iso}} n/\nu_{\text{iso}} = D_n/\nu_{\text{iso}} = 1$; a resistivity-scaling study will be presented in a forthcoming paper on benchmarking with results from other codes. We also set $n_s$ to the uniform value of $2 \times 10^{20}\ \text{m}^{-3}$, and $m_i = 1.67 \times 10^{-27}$ kg. With $RB_\phi = 4.68$ Tm at the plasma surface, the relevant Alfvén time is $\tau_A = 0.548\ \mu s$. Given the specifications from Ref. [32] being in physical units, our NIMSTELL computations for this application are in SI units. Our DESC computation for the equilibrium and geometry uses harmonics of $N_p$ to $N = 40$, though the shape of the equilibrium is described by $n = 0$ and $n = 5$. For DESC's Zernike polynomials, the radial resolution has $L = 14$, and the azimuthal resolution has $M = 12$.

Computed growth rates from scans of spatial and temporal resolution for linear computations are shown in Fig. 7. The poloidal resolution is checked with a mesh of 26 radial × 48 azimuthal elements surrounding an inner region of 144 elements, and there is packing near the $\iota = 1/2$ surface. Figure 7a shows the results from scanning the degree of the element basis functions for both magnetic representations. The horizontal axis is the polynomial degree used for the $\boldsymbol{V}$-expansion for all computations, and the continuous $\lambda_k$ polynomials for the ***H***(curl) representation for $\boldsymbol{A}$ and for $\chi$ is of one degree larger. Results for the *H*1 magnetic representation include computations with and without the stabilization terms of Eq. (13), again setting $d_d = 1$ and $d_v = 0.1$ when used. Except where noted, the *H*1 magnetic computations have $\kappa_b = 10^5$. At degree 2, the growth rate with the stabilization term is $3.1 \times 10^{-3}\ \tau_A^{-1}$ and is not shown in the figure. The computations at degrees 3 and 4 have growth rates within 0.1% of each other but are 3% larger than the computations with vector potential. A bicubic computation with the *H*1 magnetic representation and $\kappa_b$ reduced to $10^4$ (not shown) produces a growth rate within 0.7% of the vector-potential results, and the divergence error of 0.008 measured by $\int (\nabla \cdot \boldsymbol{B})^2 dVol \,/\, \int B^2 dVol$ is larger than the *H*1 results with $\kappa_b = 10^5$ by a factor of two. The scans of toroidal resolution in Fig. 7b with element bases of degree 3 and the *H*1 magnetic representation show variation when the maximum component is varied from $N = 15$ to $N = 21$ but not when increased further to $N = 31$. With $N \geq 21$, the magnetic divergence error of the *H*1 representation is approximately 100 times smaller than at $N = 15$. Thus, the diffusive correction does not appreciably enhance magnetic reconnection with $\kappa_b = 10^5$ when spatial resolution is sufficient. When scanning $\Delta t$ with adequate spatial resolution (Fig. 7c), results from all three sets of computations are similar. Empirically, the temporal convergence rate is approximately 2.

We continue two of the linear computations nonlinearly with the same physical parameters, running each through saturation of the tearing mode. Both have toroidal resolution $0 \leq n \leq 21$, given that the linear results do not change when increasing *N* to 31. One case has the *H*1 magnetic representation with all basis functions having polynomial degree 3. Because plasma pressure is extremely small, we do not include the numerical-interchange stabilization terms. The second case uses the ***H***(curl) vector potential representation with polynomial degree of 2 for all basis functions except the continuous cubic $\lambda_k$ in the ***H***(curl) expansion of $\boldsymbol{A}$ and for the scalar-potential $\chi$. Figure 8 shows the evolution of the magnetic fluctuation energies for each computation, starting from the last part of the linear computations. The Fourier components that are not part of the $n = 1, 4, 6, 9, \ldots$ linear mode family increase rapidly when the nonlinear terms are included at 1.2 ms, and the nonlinearly driven $n = 2$ component has larger energy than the $n > 1$ components of the linear tearing mode family soon thereafter. The peak $n = 1$ magnetic fluctuation energies are 0.094 J at 3.0 ms in the *H*1 magnetic-field case and 0.11 J at 3.2 ms in the lower-resolution ***H***(curl) vector-potential case. The peak energy levels agrees with JOREK's 0.1 J level that is evident in Fig. 9 of Ref. [32]. The resulting magnetic islands shown at the end of each simulation in Fig. 9 are also consistent with each other, apart from the narrow $(m = 5, n = 3)$ island in the *H*1 result. They are also consistent with Fig. 10 of Ref. [32], although a different toroidal angle is shown. Both NIMSTELL computations indicate that the magnetic fluctuation energies decay slowly after reaching their maximum amplitude. We infer that this slow decay results from the profiles adjusting to the presence of the magnetic island on a resistive-transport timescale, which is approximately 9 ms for the resistivity value used in the computations.

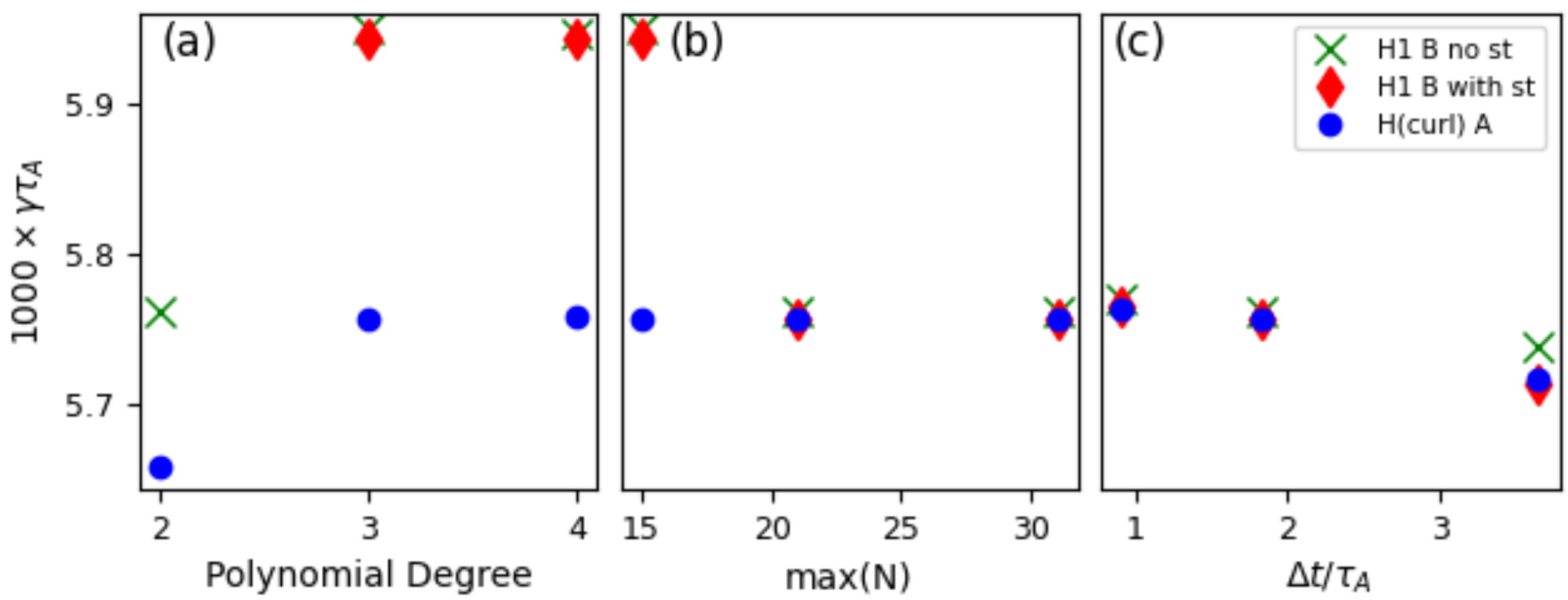

Figure 7. NIMSTELL-computed linear growth rates for the W7-A tearing computation, varying (a) the polynomial degree of the representation with $N$ from the Fourier expansion set to 15 and $\Delta t = 1.8\,\tau_a$, (b) $N$ with the polynomial degree of the $\boldsymbol{V}$-expansion set to 3 and $\Delta t = 1.8\,\tau_a$, and (c) $\Delta t$ with $N = 21$ and the degree of $\boldsymbol{V}$ set to 3. The $\lambda_k$ bases of the $\boldsymbol{H}$(curl) expansion for $\boldsymbol{A}$ is one degree larger than that of the $\boldsymbol{V}$-expansion.

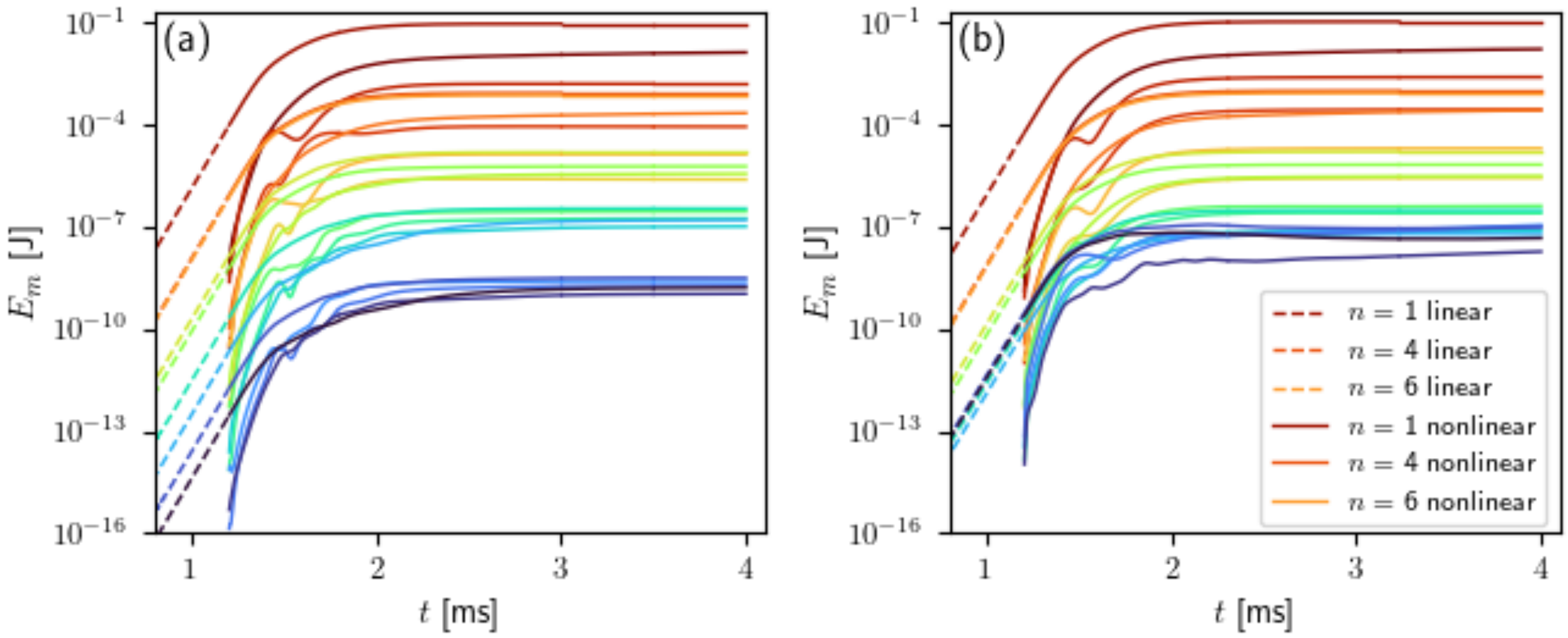

Figure 8. Evolution of magnetic fluctuation energy spectrum through nonlinear saturation for the NIMSTELL W7-A tearing computations run with (a) the $H1$ magnetic-field representation and (b) the $\boldsymbol{H}$(curl) vector-potential representation. The steady equilibrium field is not included in the traces.

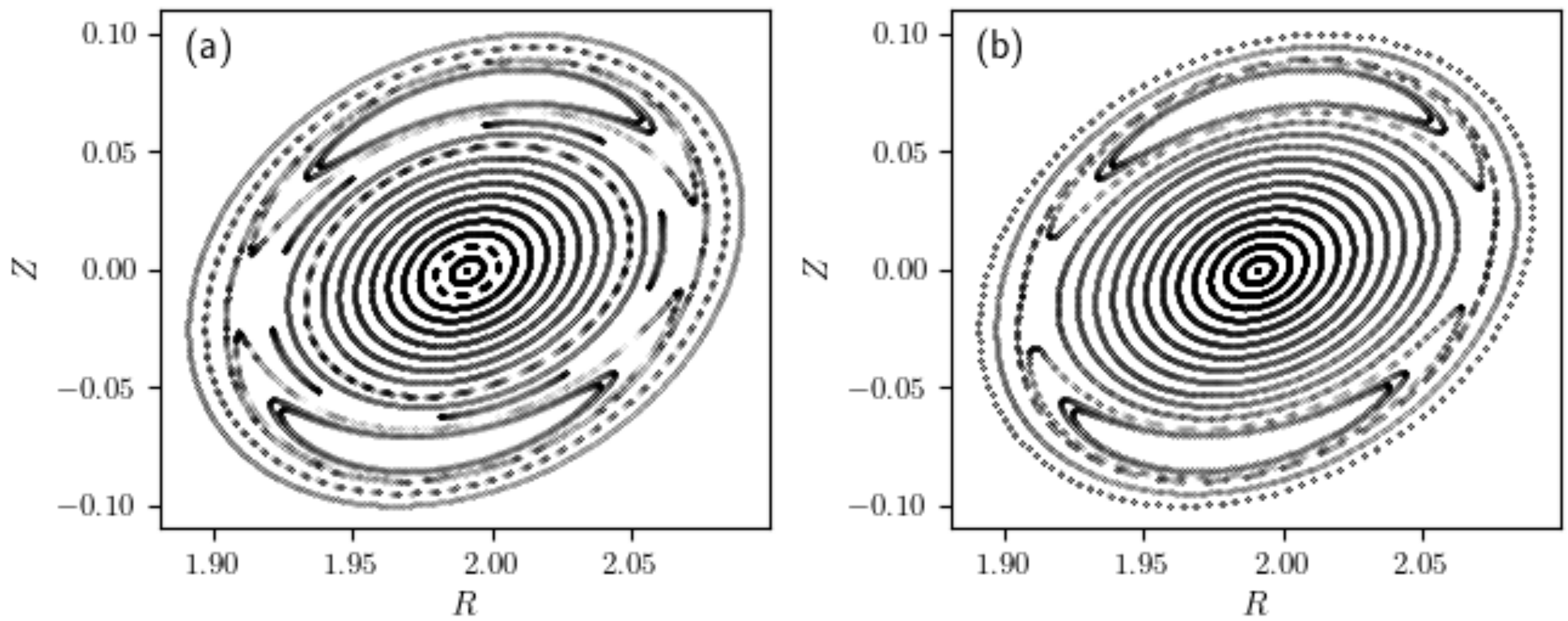

Figure 9. Magnetic Poincaré plots at $t = 4$ ms for the NIMSTELL W7-A tearing computations with (a) the *H*1 magnetic-field representation and (b) the ***H***(curl) vector-potential representation.

A final tearing-mode case demonstrates a nonlinear application of NIMSTELL for a quasi-symmetric stellarator. The geometry and profile are a modification of the optimized $N_p = 2$ quasi-axisymmetric stellarator equilibrium of Ref. [69]. Starting from the DESC team's example input for this equilibrium [70], we add the toroidal current profile $I(\psi_{tn}) = 19\psi_{tn} - 13\psi_{tn}^2 + 3.8\psi_{tn}^3$ [kA], which raises the rotational transform profile to cross the $\iota = 1/2$ surface, as shown in Fig. 10a. This change affects the fine level of quasi-symmetry described in the reference, but the errors are of order $10^{-3}$. We also add small uniform plasma pressure of 1 Pa for NIMSTELL's semi-implicit operator. The shape of the configuration is shown through the plot of the magnitude of $\boldsymbol{B}_s$ at the plasma surface in Fig. 10b. Based on $B_s{\sim}1$ T, $\rho_s = 6.35 \times 10^{-7}$ kg/m$^3$, and $R = 1.2$ m, $\tau_A = 0.78\ \mu s$. With $\eta = 1.94 \times 10^{-8}\ \Omega\text{m}$ and aspect ratio of six, the Lundquist number ($S \equiv \tau_r/\tau_A$) is $3.3 \times 10^6$. We again set $\text{Pm} = 0.001$, $\nu_\parallel/\nu_{\text{iso}} = 1000$, and and $(\Gamma - 1)\kappa_{\text{iso}} n/\nu_{\text{iso}} = D_n/\nu_{\text{iso}} = 1$. The *H*1 magnetic-field representation is used with a 34 × 48 polar mesh, packed at the $\iota = 1/2$ surface, surrounding an inner region of 144 elements. All fields are expanded in biquartic elements, toroidal resolution is $0 \le n \le 21$, and the numerical-interchange stabilization terms are not used for this case without a pressure gradient. The computation is run linearly to establish the tearing mode, and nonlinear terms are activated at 50 ms. As in the W7-A computation, Fourier components that are not part of the linear mode family are driven rapidly after the nonlinear terms are included (Fig. 11a). The width of the saturated magnetic island (Fig. 11b) is significant with respect to the minor radius of the stellarator.[2]

[2] We remind readers that the initial state is a modification of the optimized result of Ref. [69], where the $\iota$-profile does not cross 1/2.

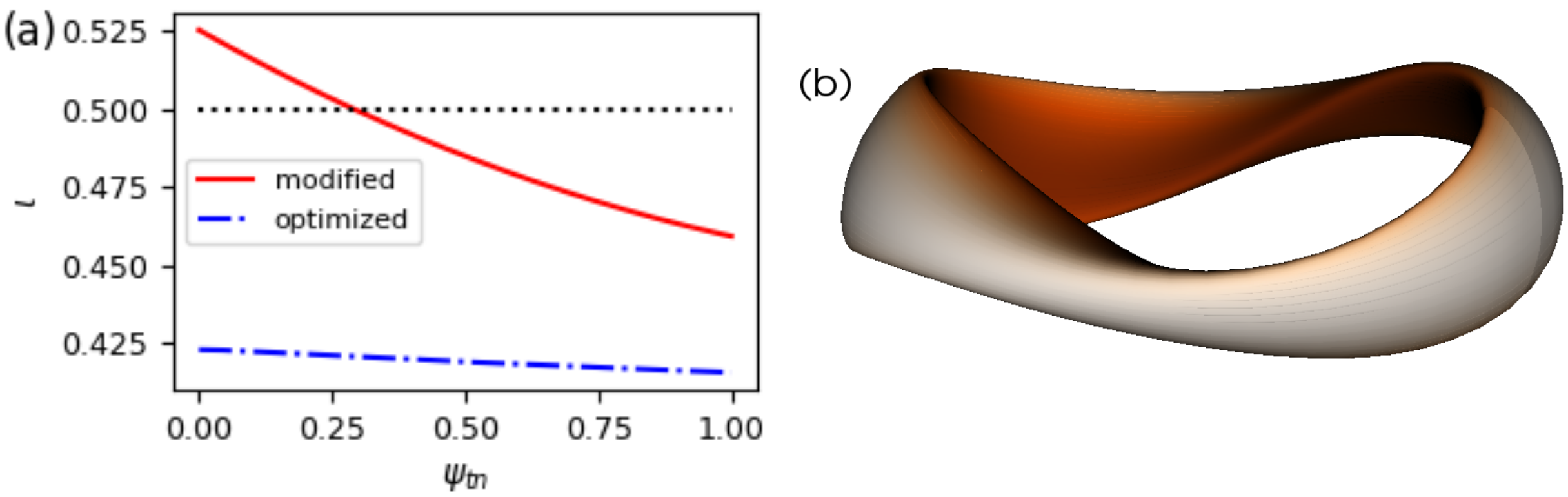

Figure 10. Optimized and modified rotational transform profiles (a) and $|\boldsymbol{B}_s|$ at the plasma surface (b) for the quasi-axisymmetric NIMSTELL tearing computation.

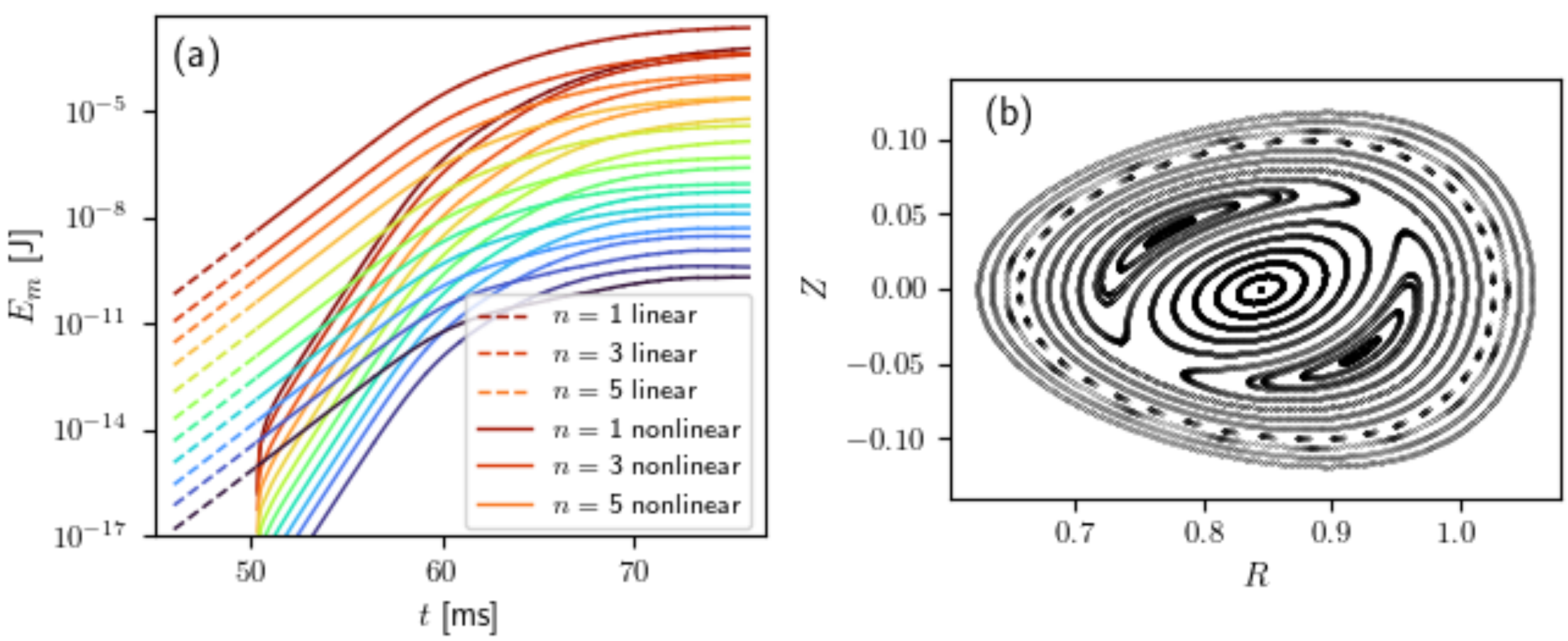

Figure 11. Nonlinear results on the quasi-axisymmetric tearing computation showing (a) evolution of magnetic fluctuation energies and (b) magnetic topology via Poincaré surface of section for $t = 76\ \mathrm{ms}$.

### 4.3 Anisotropic thermal conduction

To demonstrate numerical convergence for conditions of anisotropic transport, we consider anisotropic thermal conduction with fixed magnetic field without flow. We first have NIMSTELL compute a helically symmetric current-free magnetic field as a magnetostatics problem in a straight periodic geometry, and then use it to solve a time-dependent heat transport computation until steady conditions are reached. Although an analytical solution with anisotropic thermal conduction is not known for this configuration, the computed central temperature provides a measure of the artificial perpendicular heat transport at a given resolution when using a large $\kappa_\parallel/\kappa_\perp$-ratio [34].

The analytical form of the magnetic configuration is defined by the magnetic potential $\Phi_m$ with $\boldsymbol{B} = \nabla\Phi_m$. The single-mode solution to $\nabla^2\Phi_m = 0$, plus a secular axial component, is

$$\Phi_m(r,\theta,z) = C I_l\left(\frac{N_p r}{R}\right)\cos\left(l\theta - \frac{N_p z}{R}\right) + z\ ,$$

where $I_l(x)$ is the $l$-th order modified Bessel function of the first kind, $N_p$ is the axial-period number for an equivalent major radius of $R$, and $C$ is a constant that sets the magnitude of the field relative to the normalized secular component. We use a configuration with $l = 2$, $R/N_p = 4$, and $C$ is set so that the amplitude of $B_r$ is 0.1 at $r = 1$. The closed magnetic-flux region is elongated in the $r$-$\theta$ plane with a separatrix having $x$-points at $r = 3.46$. The magnetostatics problem determines $\boldsymbol{A}$ and $\chi$ from a modified version of the implemented Eq. (12) such that $\nabla\times\nabla\times\boldsymbol{A} + C_d\boldsymbol{A} + \nabla\chi = 0$ with $\nabla^2\chi = \nabla\cdot\boldsymbol{A}$ in the interior of the domain. The parameter $C_d$ is for numerical solvability and is set to $10^{-6}$. Boundary conditions impose $\hat{\boldsymbol{n}}\times(\nabla\times\boldsymbol{A} - \nabla\Phi_m) = 0$ and $\chi = 0$ with $\hat{\boldsymbol{n}}$ being the surface normal. The heat transport problem solves Eqs. (3) and (4) with $\boldsymbol{V} = 0$, $Q = 4$, $\kappa_\perp = 1$, $\kappa_\parallel = 10^6$, and $T = 0$ along the boundary. Rapid parallel transport maintains the open-field temperature at its boundary value. Numerically, the polynomial degree of $\lambda_i$ for the ***H***(curl) $\boldsymbol{A}$ and $\chi$ is one larger than that used for $T$.

We consider three sets of computations. One has a cylindrical domain of radius 4, and the other two have elliptical domains with semi-major axis of 4 and ellipticity of 0.96 that are aligned with the helical closed-flux region. Of the two shaped computations, one has its plane of elements twist with the closed-flux region. The geometry and magnetic potential in this set are fully represented by the $n = 0$ and $n = 1$ components of the Fourier expansion for the axial coordinate. An entire $2\pi$ rotation of the elliptical shape is required, so the configuration has $N_p = 2$ and $R = 8$. With this domain, the heat transport problem is 2D, and the temperature is fully computed in NIMSTELL's $n = 0$ component. The other set of shaped computations maintains the same azimuthal orientation of the elements within the poloidal plane but distorts their poloidal positions to follow the elliptical shaping. These computations represent a single axial period with $N_p = 1$ and $R = 4$, and the geometry, magnetic potential, and temperature have broad axial spectra. The two types of shaped meshing are illustrated in Fig. 12. None of the three sets of computations has a mesh that is precisely aligned with the magnetic flux surfaces, but the elliptically shaped meshes have better alignment than the circular mesh. The center of each mesh transitions to a distorted rectangle, as used for all MHD computations reported above.

Computations with the twisting elliptical mesh have two Fourier components, *i.e.* $N = 1$, and converge with increasing resolution of the $r$-$\theta$ plane. The results in Fig. 13a show an *h*-convergence series with bicubic elements for $T$ with the number of elements in the radial and azimuthal directions varied simultaneously. It also shows a *p*-convergence series with a 28×32 annular mesh surrounding the central region of 64 elements and varied polynomial degree. The mesh is not tailored to match the shape of the separatrix, and the *h*-series produces an empirical convergence rate of 2, due to the boundary layer at the separatrix. Nonetheless, the results are accurate even with relatively coarse resolution. The results shown in Fig. 13b for the cylindrical and elliptically shaped non-twisting configurations have biquintic elements for $T$ and a 31×40 annular mesh surrounding a central region of 100 elements. Resolution is only varied for the Fourier representation over the axial coordinate. Overall, the results are far less accurate than those obtained with the twisting mesh because the Fourier expansion must reproduce the axial variation of $\boldsymbol{A}$ and $T$. Elliptical shaping improves accuracy at low resolution, relative to using the cylindrical mesh, but it is not as effective as twisting the mesh to reduce the dimensionality of the problem.

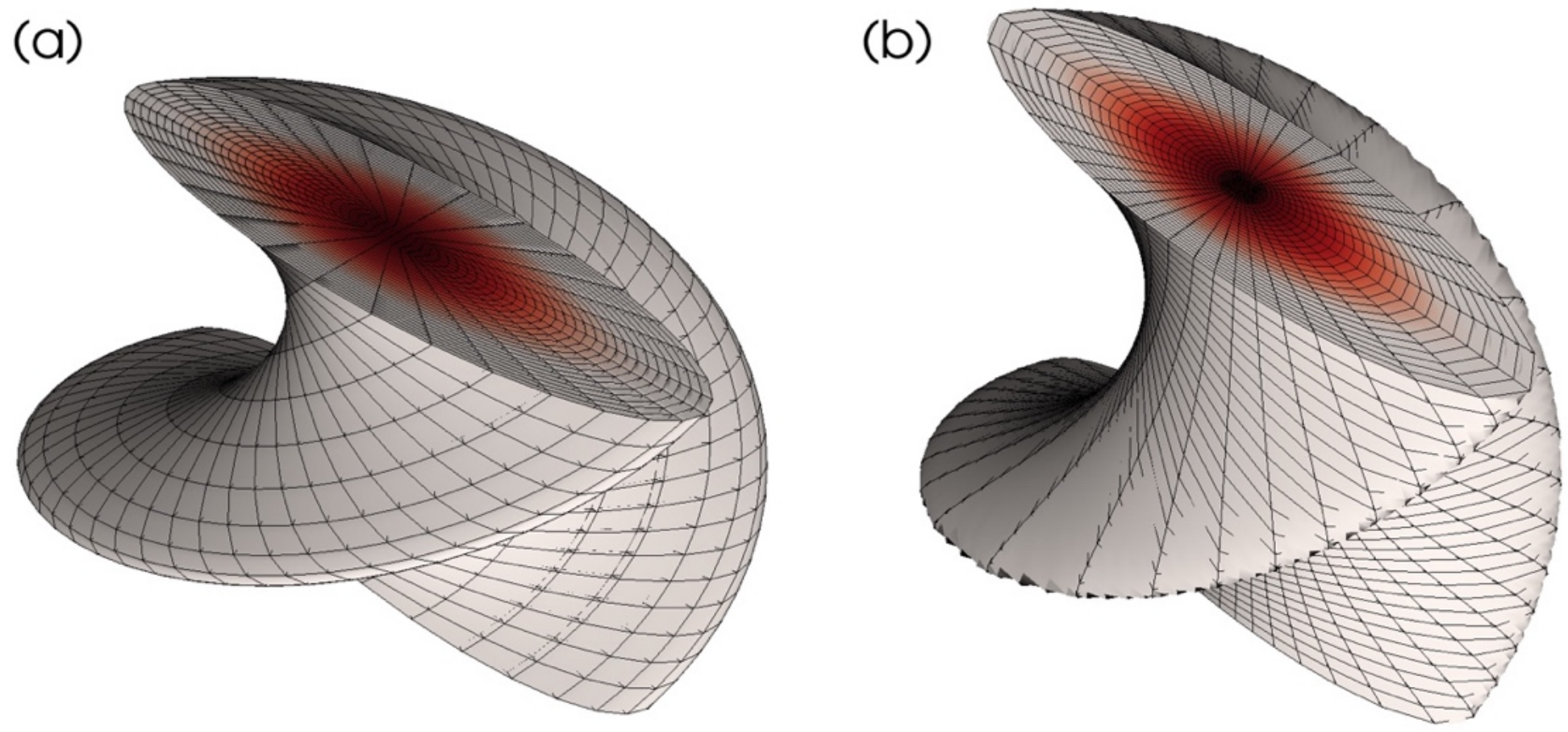

Figure 12. Plots of meshing and computed temperature for the anisotropic heat transport problem. The twisting type of mesh is shown in (a), and the distorting mesh is shown in (b). The plotted mesh lines do not show the curvature of element sides. The axial divisions are for plotting purposes only, and half of the $N_p = 2$ domain is plotted in (a). The major axis of the elliptical mesh is aligned with the orientation of the *x*-points.

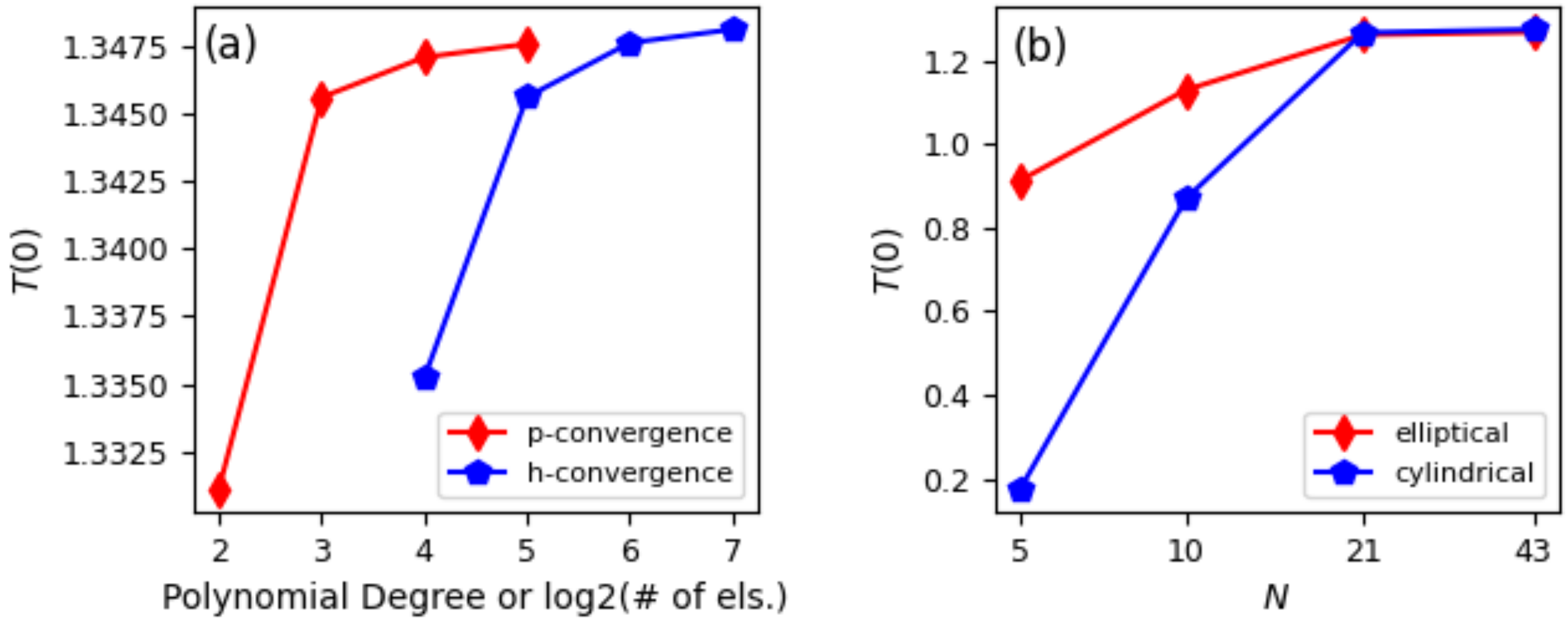

Figure 13. Results on central temperature for the heat transport problem, as computed with the (a) twisting elliptical mesh and (b) without twist for the elliptically shaped and circular meshes. The two curves in (a) show convergence with respect to polynomial degree ("p") for 960 elements and with respect to mesh resolution ("h") for bicubic basis functions for $T$. For the *h*-series, the horizontal axis shows the base-2 logarithm of the number of elements in the azimuthal direction. Poloidal resolution is held fixed in (b) as resolution over the axial direction is varied.

## 5. Discussion and conclusions

When using the *H*1 magnetic representation, the NIMSTELL numerical methods follow those used in NIMROD. Nonetheless, the modifications from the NIMROD implementation are extensive. The physical differential operations in the poloidal plane and the Jacobian of the mapping are independent of the Fourier projections in NIMROD. Derivatives are fully formed in the Fourier representation, and nonlinear products are formed separately through fast Fourier transforms (FFTs) to a mesh over the toroidal angle. In contrast, forming differential operations in NIMSTELL involves partial derivatives with respect to the element coordinates $\xi$ and $\eta$ and the generalized toroidal angle $\zeta$ in the Fourier representation that are transformed to the $\zeta$-mesh, where differentiation is completed through products with the reciprocal basis vectors $\nabla u_i$. The reciprocal basis vectors and Jacobian are formed on the $\zeta$-mesh because it is the coordinates $\boldsymbol{R}(\xi, \eta, \zeta)$ that have Fourier expansions. The operations on the $\zeta$-mesh account for the reciprocal bases needed for the test functions, in addition to those for the basis functions, before applying FFTs back to the Fourier representation.

Another important extension for practical application of NIMSTELL is having preconditioning operations include multiple Fourier components according to the symmetry of a configuration. The coupling of Fourier components from the shaping of modern stellarators is too significant to rely on iteration without Fourier coupling in the preconditioning, as is typically used for nonlinear NIMROD computations. Nonetheless, covering complete mode families may be infeasible in large production problems. For NIMROD and NIMSTELL, we typically use the SuperLU_DIST sparse direct solver [65] for our preconditioning operations. This has proven effective and practical for NIMROD and smaller NIMSTELL applications. However, preconditioning matrices can be too large to factor, despite our recent memory-related optimization, such as using blocked symbolic operations. An example is the vector-potential computation with $N = 31$ in Fig. 7b. Whereas the memory of three of the National Energy Research Scientific Computing Center's Perlmutter CPU nodes (512 GB for two AMD EPYC 7763 CPUs [71]) accommodates the preconditioning matrices covering the three mode families for all fields in the magnetic-component computation with $N = 31$, that is not the case for the $N = 31$ vector-potential computation. For that computation, we split each mode family into two bands with overlap of 2 Fourier components, which increases the iteration count for the $\boldsymbol{V}$-advance from two to hundreds, even with a loosened tolerance, and the computational time per step is more than 50 times larger than the *H*1 computation. For large problems, it may be more effective to include all Fourier components in mode families using incomplete factorization, which remains to be explored.

The option to use the ***H***(curl) vector representation is distinct for NIMSTELL. The aim of avoiding magnetic divergence error altogether prompted the analysis shown in Appendix A. Its results on the tearing-mode tests of Section 4.2 agree quantitatively with those of the *H*1 magnetic representation linearly and nonlinearly. Because the ***H***(curl) implementation uses polynomial basis functions that are of one degree larger than those for the other fields and simultaneously solves a scalar for the gauge condition, there is an associated computational cost. For the $N = 21$ linear computations of Fig. 7b, the computational time per step of the ***H***(curl) computation is 80% larger than that of the *H*1 computation without stabilization terms and 50% larger than the one with stabilization terms. It is possible that the damping method for the gauge can be applied with a time-split advance, thereby reducing the algebraic system size for $\Delta\boldsymbol{A}$. Another computational point is that our semi-implicit method uses $\nabla\boldsymbol{B}$ from the evolving fields in nonlinear computations, so our nonlinear vector-potential computations perform an extra *H*1 projection of magnetic components at each step. The greater concern, raised in Section 4.1, for the NIMSTELL vector-potential representation is the minimum level of resistivity needed for a given mesh when plasma pressure is important. A finite-element stabilization for the vector-potential advance [72] may help, though application to magnetic-confinement should avoid artificial magnetic reconnection, which we also require of our divergence cleaning with magnetic representations. Investigation of vector-potential stabilization methods for NIMSTELL is left for future work.

Use of the generalized toroidal angle in NIMSTELL is largely unexplored at this time, meaning that we have taken $\zeta = \phi$ in almost all computations, including those presented here. We foresee optimization of the mapping to achieve distinct but likely overlapping benefits of improving numerical convergence and minimizing Fourier coupling in the algebraic systems. Depending on the application, optimization may involve a combination of rotating the plane of elements, as considered in our thermal-conduction tests of Section 4.3, and modifying how $\phi$ is related to $\zeta$.

A more foundational aspect of preprocessing for stellarator applications is that NIMSTELL and other stellarator MHD codes rely on external equilibrium solutions. For NIMSTELL, that is a separate computational step that is performed with DESC [48], and equilibria previously computed with VMEC [73] can be handled through DESC's ability to read VMEC output and re-solve the force balance. For NIMSTELL computations that retain the interpolated external equilibrium in the steady-state fields, it is important for those fields to satisfy force-balance accurately. That accuracy depends on both the numerical quality of the external equilibrium computations and on the resolution of the NIMSTELL representation.

The computations presented in Section 4 show that the *H*1 magnetic representation with diffusive correction of magnetic divergence error is suitable for stellarator applications. Although the Fourier representation does not allow direct application of $\boldsymbol{B} \cdot \hat{\boldsymbol{n}} = 0$ boundary conditions through essential conditions on the *H*1 space, implicit damping of $\boldsymbol{B} \cdot \hat{\boldsymbol{n}}$ through surface integrals is effective and does not adversely affect the conditioning of the algebraic system for $\Delta\boldsymbol{B}$. The semi-implicit operator computed from 3D fields stabilizes the temporally staggered leapfrog method in strongly shaped configurations and allows separate advances of each physical field. Also, the numerical-interchange stabilization method from our earlier NIMROD work yields convergence from the stable side in the presence of bad curvature. Although only confirmed here with cylindrical tests, results on pressure-driven MHD modes in stellarator configurations will be presented in a forthcoming benchmarking article. Including multiple Fourier components in our preconditioning operations, according to the symmetry of a given application, has proven to be important for reducing Krylov-space iteration to a practical level. As for the vector-potential option, it requires further numerical development to be used reliably in all applications. It is possible that solutions to the apparent coercivity problem at small resistivity values already exist and will be explored in future work, along with further solver and mapping improvements.

## 6. Acknowledgments

The first author is pleased to acknowledge discussions with former students Dr. Brian S. Cornille and Mr. Colin M. Guilbault that contributed to early aspects of the NIMSTELL effort. Funding for this work has been provided by U.S. Dept. of Energy awards DE-SC0018642 and DE-FG02-99ER54546. Award DE-SC0024548 for the HiFiStell Project also funds this work, and prior to April 1, 2025, Princeton University provided subcontract SUB0000763 for the University of Wisconsin-Madison effort.

This research uses resources of the National Energy Research Scientific Computing Center (NERSC), a Department of Energy User Facility, NERSC award FES-ERCAP 32189. Our research also uses computational resources at the University of Wisconsin-Madison's Center for High Throughput Computing (2006), doi:10.21231/GNT1-HW21.

## 7. Appendix A. Analysis of MHD representation with vector potential

The performance of different numerical formulations for inhomogeneous plasma profiles cannot be fully described through analytics. Thus, we use a numerical eigenvalue solver with a flexible numerical representation over the radial coordinate while assuming $\exp(im\theta + ikz)$ over the azimuthal and axial directions for a chosen $(m, k)$ component. This code had previously been used to develop the numerical interchange stabilization method described in Section 3.3. The code allows rapid investigation of different formulations for meaningful test cases. Here, we use it to investigate the ***H***(curl) vector-potential representation with various formulations for setting the gauge condition and to compare the influence of changing the degree of the polynomial basis functions for potentials relative to those for the flow velocity and the scalar pressure. Because the element expansion is only over one coordinate and a time-advance algorithm is not represented, the eigenvalue computations provide relevant but incomplete information for NIMSTELL. The algebraic eigenvalue problems resulting with the different numerical formulations are solved with appropriate routines from the LAPACK library [74].

Using vector potential in the linear ideal-MHD system without equilibrium flow, and assuming time-dependence of the form $\exp(-i\omega t)$, leads to the eigenvalue problem,

$$-i\omega\rho_0 \boldsymbol{V} = \nabla\cdot\left[\frac{1}{\mu_0}\boldsymbol{B}_0(\nabla\times\boldsymbol{A}) + \frac{1}{\mu_0}(\nabla\times\boldsymbol{A})\boldsymbol{B}_0 - \underline{\mathbf{I}}\left(\frac{1}{\mu_0}\boldsymbol{B}_0\cdot\nabla\times\boldsymbol{A} + p\right)\right], \tag{A1}$$

$$-i\omega p = -\boldsymbol{V}\cdot\nabla p_0 - \Gamma p_0 \nabla\cdot\boldsymbol{V}\text{, and} \tag{A2}$$

$$-i\omega\boldsymbol{A} = \boldsymbol{V}\times\boldsymbol{B}_0 - \nabla\chi \tag{A3}$$

with boundary conditions $rA_\theta = A_z = V_r = 0$ for an ideal conductor at $r = a$. The equilibrium fields are indicated by the "0" subscripts. The first-order form of the system represents how the equations are solved in time-dependent computations, and ideal MHD is the numerically challenging limit of vanishing dissipation. An alternative system uses the total pressure $p_T$ instead of the sum of separate plasma and magnetic pressures in (A1), and (A2) is replaced by the equation for the total pressure,

$$-i\omega p_T = \boldsymbol{B}_0\cdot\nabla\boldsymbol{V}\cdot\boldsymbol{B}_0 - \left(\frac{B_0^2}{\mu_0} + \Gamma p_0\right)\nabla\cdot\boldsymbol{V} - \boldsymbol{V}\cdot\nabla\left(\frac{B_0^2}{2\mu_0} + p_0\right). \tag{A4}$$

When using ***H***(curl) representations for $\boldsymbol{A}$, the polynomial expansion for $A_r$ is discontinuous in $r$ and of one degree smaller than those for $rA_\theta$ and $A_z$, which are continuous. We also consider computations that use an *H*1 representation for all components of $\boldsymbol{A}$.

Three different gauge formulations have been investigated with the eigenvalue computations. The first is the Weyl gauge with $\chi = 0$ and ***H***(curl) representations for $\boldsymbol{A}$. This formulation would reduce the computational cost of time-dependent 3D simulations by avoiding the extra field $\chi$. The second formulation is described by $\nabla^2\chi = d_A\nabla\cdot\boldsymbol{A}$ . It leads to the Coulomb gauge by damping $\nabla\cdot\boldsymbol{A}$. This property is shown by taking the divergence of the ideal part of (A3) and replacing $\nabla^2\chi$ with $d_A\nabla\cdot\boldsymbol{A}$ to produce $-i\omega\nabla\cdot\boldsymbol{A} = \nabla\cdot(\boldsymbol{V}\times\boldsymbol{B}_0) - d_A\nabla\cdot\boldsymbol{A}$. For $\boldsymbol{A}$ in the ***H***(curl) space, this formulation requires integrating the weak form of the $\chi$-equation by parts. A third formulation also leads to the Coulomb gauge, but it diffuses $\nabla\cdot\boldsymbol{A}$ by setting $\chi = -D_A\nabla\cdot\boldsymbol{A}$. Repeating the substitutions produces $-i\omega\nabla\cdot\boldsymbol{A} = \nabla\cdot(\boldsymbol{V}\times\boldsymbol{B}) + \nabla^2 D_A\nabla\cdot\boldsymbol{A}$. Here, the additional condition $\oint_{\partial\Omega} d\boldsymbol{S}\cdot\boldsymbol{A} = 0$ is needed to remove numerical monopoles in $\boldsymbol{A}$. This formulation may be used with $\boldsymbol{A}$ in ***H***(curl), provided that the $\nabla\cdot\boldsymbol{A}$ projection is integrated by parts. Alternatively, if $\boldsymbol{A}$ is in *H*1 and the $\nabla\cdot\boldsymbol{A}$ projection is not integrated by parts but the $\nabla\chi$ projection is, then $\chi$ may be of the $L^2$ space. This would be convenient for eliminating the coefficients of $\chi$ in terms of those for $\boldsymbol{A}$ when solving the algebraic system. Taking the limit of large $d_A$ or $D_A$ enforces the Coulomb gauge, strictly, in the damping and diffusive formulations.

Basic numerical properties of these three formulations are evident from eigenvalue results for equilibria with uniform $\boldsymbol{B}_0$ in the axial direction and with uniform $p_0$ and $\rho_0$. The analytical spectrum of this system includes a degenerate torsional Alfvén spectrum, the sound wave as the accumulation point of slow modes, and the fast Sturmian modes [13]. We choose a normalized configuration with $B_{0z}$, $\rho_0$, $\Gamma$, cylinder radius $a$, and $\mu_0$ set to unity and $p_0 = 1/2$. For $m = 1$, $k = 2$, for example, the Alfvén spectrum is at $|\omega| = 2$, the first two fast modes are at $|\omega| = 3.088$ and $6.875$, the sound wave is at $|\omega| = \sqrt{4/3}$, and the largest-frequency slow mode is at $|\omega| = 1.245$. There are no modes below the sound accumulation point, between the highest-frequency slow wave and the Alfvén spectrum, and between the Alfvén spectrum and the first fast mode. A comparison of computed $|Re\{\omega\}|$ for the three gauge formulations with either plasma or total pressure is shown in Fig. 14a. Three elements are used over $r$ for each computation, and $p$, $V_r$, $V_\theta$, and $V_z$ are cubic within each element and continuous across element interfaces for their *H*1 space. For the Weyl and damping formulations, $\boldsymbol{A}$ is in ***H***(curl) with cubic $A_r$ and quartic $rA_\theta$ and $A_z$, and the potential $\chi$ for the damping formulation is quartic and in *H*1. For the dissipative formulation, we show results where $A_r$, $A_\theta$, and $A_z$ are in *H*1 and quartic, $\chi$ is in *H*1 and cubic, and the $\nabla \cdot \boldsymbol{A}$ projection is integrated by parts.

For all three gauge formulations, results computed with total pressure instead of plasma pressure reproduce the analytical frequencies, accurately. Computations with plasma pressure produce two numerical modes between the Alfvén spectrum and the first fast mode. The computed eigenfunctions (not shown) indicate that these numerical modes are noisy in *r* and have relatively little coupling to $\boldsymbol{A}$. For the Weyl and damping formulations with $\boldsymbol{A}$ in ***H***(curl), there are 16 numerical modes having $Re\{\omega\} = 0$. There is no damping of these non-oscillatory modes in the Weyl formulation. With the damping formulation to control $\nabla \cdot \boldsymbol{A}$, 11 of these modes have $Im\{\omega\} = -100$, which is the applied $d_A$-value. The number 11 corresponds with the degrees of freedom in the quartic *H*1 expansion of $\chi$ after considering its boundary condition and the regularity condition at $r = 0$. The remaining 5 non-oscillatory modes with the damping formulation have eigenfunctions where the amplitudes of $p$ and $\boldsymbol{V}$ are near roundoff. These eigenfunctions are solenoidal perturbations of $\boldsymbol{A}$ at the mesh scale that do not couple to the fluid and are not damped in the absence of resistive dissipation.

The diffusive formulation with $D_A = 100$, $\chi$ in *H*1, and the projection of $\nabla \cdot \boldsymbol{A}$ integrated by parts is also accurate for the uniform equilibrium, regardless of whether $\boldsymbol{A}$ is in *H*1 (shown) or in ***H***(curl) (not shown). With $\boldsymbol{A}$ in *H*1, there is one more non-oscillatory mode. In contrast, computations with $\boldsymbol{A}$ in *H*1, $\chi$ in $L^2$, and projections of $\nabla \cdot \boldsymbol{A}$ not integrated by parts fail to produce an Alfvén spectrum. Moreover, the imaginary parts indicate growing modes, and a numerical MHD formulation that produces growing modes for a spatially uniform equilibrium is not acceptable. This last formulation may require the space of $\boldsymbol{A}$ and $\chi$ functions to satisfy the divergence-stability condition [75].

Using expansions of $\boldsymbol{A}$ with the continuous $\lambda_k$ polynomials of the same degree as those for $p$ and $\boldsymbol{V}$ would reduce computational costs in time-dependent applications. Thus, we also consider results with cubic expansions for all fields except $A_r$, which is quadratic. Figure 2b shows that regardless of whether plasma or total pressure is used, the ***H***(curl) computations with damping have oscillatory modes below the sound wave and between the slow and Alfvén modes. There are fewer non-oscillatory modes, but one of these modes has $Im\{\omega\} > 0$ when the system includes plasma pressure. The computation with $\boldsymbol{A}$ in *H*1, the diffusive gauge formulation, and total pressure has an accurate real spectrum, apart from the non-oscillatory modes.

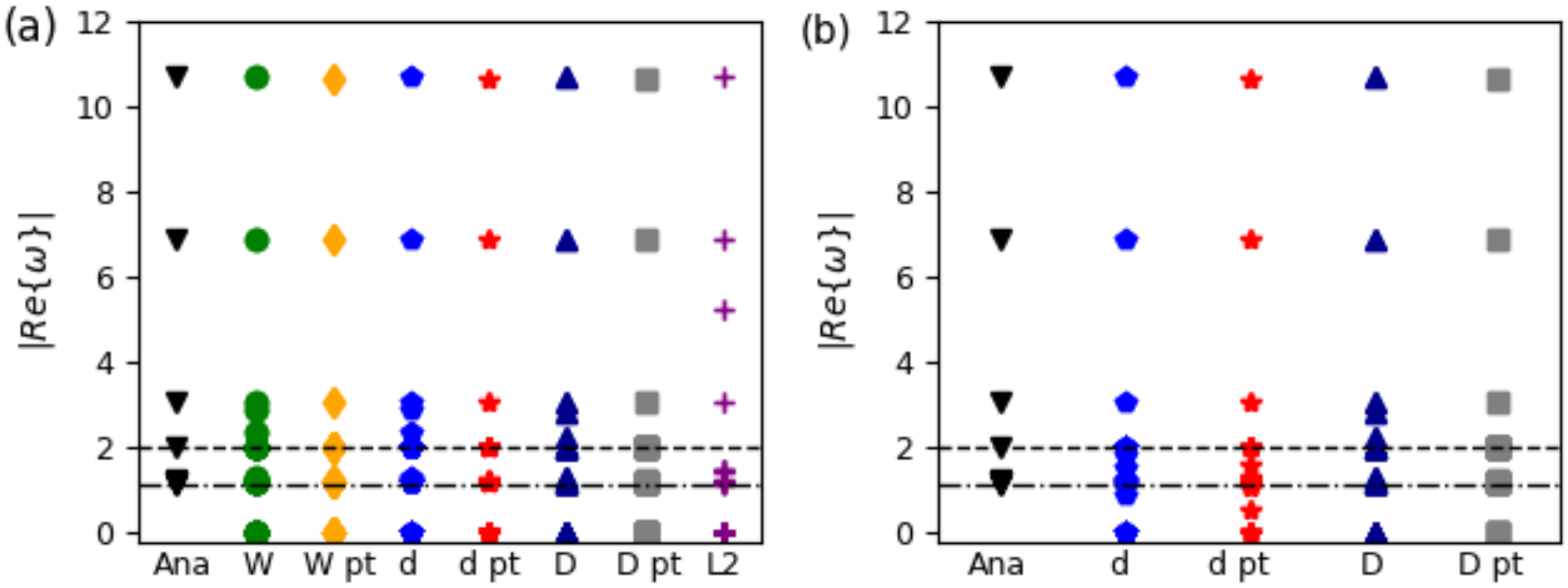

Figure 14. Two comparisons of $|Re\{\omega\}|$ from the eigenvalue computations with different gauge and pressure formulations for the uniform equilibrium and $m = 1, k = 2$. Results for $\boldsymbol{A}$ of one polynomial degree greater than $p$ and $\boldsymbol{V}$ are shown in (a), and results for $rA_\theta$ and $A_z$ of the same degree as $p$ and $\boldsymbol{V}$ are shown in (b). The column labeled "Ana" shows reference analytical values. The computational results are for the Weyl formulation ("W"), damped formulation with $\boldsymbol{A}$ in $\boldsymbol{H}$(curl) ("d"), dissipative formulation with $\boldsymbol{A}$ and $\chi$ in $H1$ ("D"), and dissipative formulation with $\chi$ in $L^2$ ("L2"). Labels with "pt" indicate the use of total pressure from Eq. (A4). The dash-dot lines show the slow-mode accumulation point, and the dashed lines show the Alfvén spectrum.

Other information that can be obtained from the ideal-MHD eigenvalue computations is whether a formulation converges on local interchange from growth rates that are too small ("stable side") or too large ("unstable side"). The preference in early ideal-MHD eigenvalue computations was to converge from the unstable side to avoid missing growing modes [2]. However, Lütjens and Luciani note that time-dependent nonlinear computations need to avoid the growth of unresolved or analytically stable modes [62]. This prompted the development of the specific stabilization scheme for NIMROD, described in Section 3.3, to make it converge on resonant interchange from the stable side.

We again consider the cylindrical profiles described in Section 4.1 and $m = 4$ modes with $k = -1.5$ and $k = -1.784$, which are resonant at $r = 0.507$ and $r = 0.371$ with $D_s$ parameters of 0.200 and 0.443, respectively. Results on the fastest-growing mode from computations with the Weyl and damping formulations for the analytically unstable $k = -1.784$ wavenumber are shown in Fig. 15a. These computations have $\boldsymbol{A}$ in $\boldsymbol{H}$(curl) with quartic $rA_\theta$ and $A_z$ and cubic $H1$ expansions for $p$ and $\boldsymbol{V}$. The elements have uniform spacing over $r$ to demonstrate convergence without prior knowledge of localized behavior for a particular mode. Apart from the 20-element computations, where the damping formulation produces a weakly growing mode and the Weyl computation does not, the growth rates from the two sets of computations agree to within 0.001%. Growth rates from the 160- and 320-element computations agree to within 0.23%. Part of the eigenfunction for the unstable mode from the 80-element computation with damping is plotted in Fig. 15b to show the extent of radial localization. The phase of the mode with respect to $\exp(im\theta + ikz)$ is arbitrary. The observed convergence from the stable side when using $\boldsymbol{A}$ of greater polynomial degree than $p$ and $\boldsymbol{V}$ is consistent with the analysis in Ref. [62] for their lower-order representations of displacement and magnetic flux with reduced-MHD.

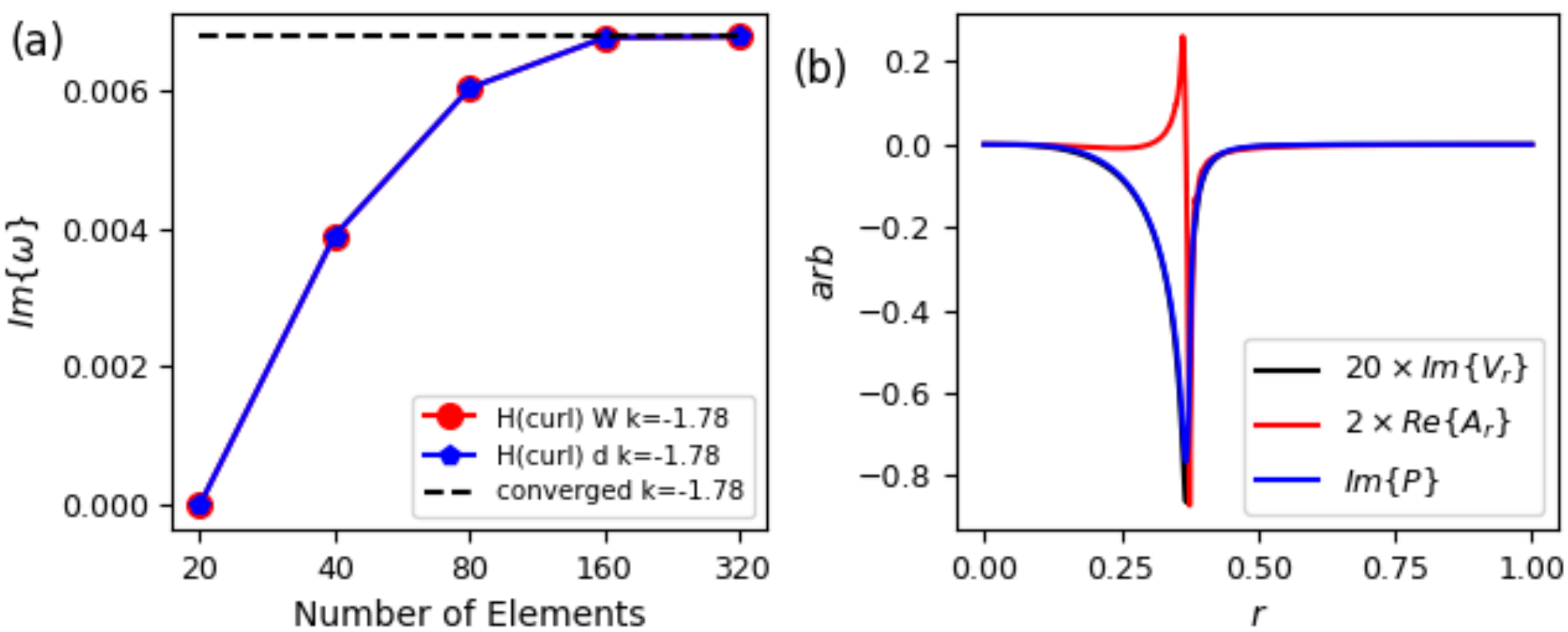

Figure 15. Results for the $m = 4$, $k = -1.784$ unstable interchange mode that is resonant at $r = 0.371$ . Frame (a) shows numerical convergence on growth rates for the Weyl and damping formulations with $d_A = 1$ for the latter. Frame (b) shows eigenfunction components of the 80-element damping computation.

Figure 16 shows results for the analytically stable $k = -1.5$ wavenumber from four sets of computations. The Weyl formulation does not produce a growing mode in the computations with 160 or fewer elements. The damping formulation with $\boldsymbol{A}$ in ***H***(curl) and $d_A = 1$ produces weakly growing modes at all resolution levels, and the 320-element growth rate is slightly faster than the one from the Weyl formulation. Other formulations do not perform as well. Computations with the diffusive formulation, $D_A = 1$, and $\boldsymbol{A}$ in ***H***(curl) produce modes that grow faster than the damping formulation. When the *H*1 expansion is used for $\boldsymbol{A}$, the computations produce overstable modes that grow very rapidly. Other formulations also perform poorly. Computations with damping and $\boldsymbol{A}$ in ***H***(curl) but of the same polynomial degree as $p$ and $\boldsymbol{V}$ produce a non-oscillatory mode with $Im\{\omega\} = 0.14$. Computations with total pressure instead of plasma pressure also find growing modes with $Im\{\omega\}$ of order 0.1. Computations with $\boldsymbol{A}$ in *H*1, $\chi$ in $L^2$, and the $\nabla \cdot \boldsymbol{A}$ projection not integrated by parts produce overstable modes with $Im\{\omega\}$ of order 1. These results show that none of the formulations reproduce a strictly stable spectrum when the conditions are close to marginal stability, but the Weyl and damping formulations with $\boldsymbol{A}$ in ***H***(curl) and $\lambda_k$ of one degree greater than the bases for $p$ and $\boldsymbol{V}$ predict the weakest numerical growth. The equilibrium profile has $D_s \to \infty$ for $r \to 0$, which makes these computations challenging for all formulations. We have also found that adding the interchange stabilization terms to eigenmode computations with vector potential does not improve the numerical behavior.

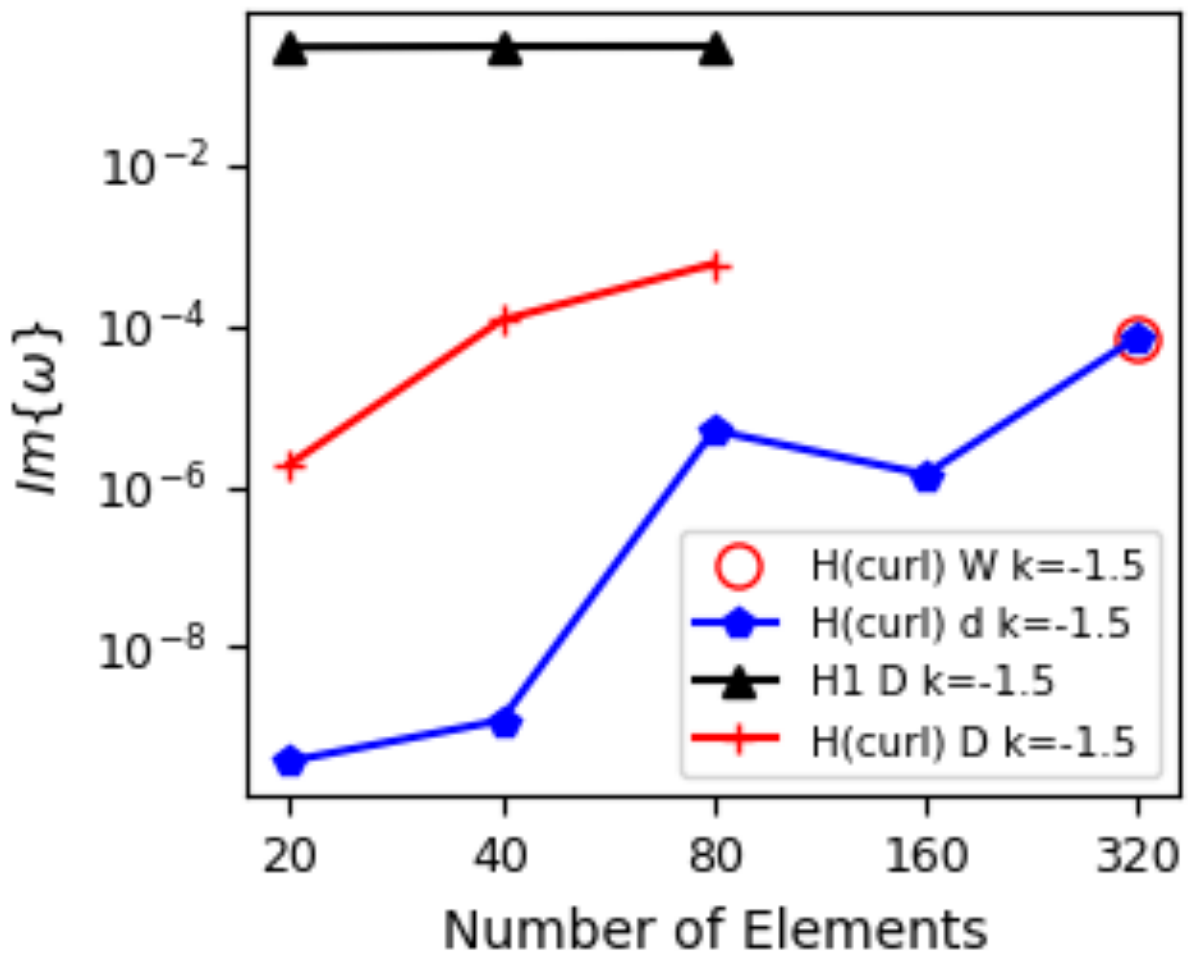

Figure 16. Results for the $m = 4$, $k = -1.5$ stable interchange mode that is resonant at $r = 0.443$. The symbols and traces show results with the Weyl, damping, and diffusive formulations with $\boldsymbol{A}$ in the spaces indicated in the figure legend.

Because NIMSTELL is intended to allow *p*-refinement, we also consider sets of 40-element computations that scan the polynomial degree of the basis functions. They have $\boldsymbol{A}$ in ***H***(curl) with the continuous $\lambda_k$ of one degree larger than the *H*1 expansions for $p$ and $\boldsymbol{V}$, and the number of Gaussian integration points in each element is increased with the polynomial degree to ensure accurate projections. The computations use the damping formulation with $d_A = 1$. Although *p*-refinement is not recommended for modeling localized behavior, Fig. 17 shows that accurate results can be obtained for the unstable $k = -1.784$ wavenumber, and the computations avoid growing solutions for the analytically stable $k = -1.5$ wavenumber to a greater extent than with *h*-refinement. For comparison, the unstable-mode computation with quintic polynomials has fewer degrees of freedom than the 80-element computation with cubic polynomials, and the quintic computation is slightly more accurate.

For the reasons discussed in Section 4.1, the NIMSTELL code allows the user to choose either the new ***H***(curl) expansion for the vector potential or the *H*1 expansion of magnetic-field components. The latter is used in NIMROD, and for NIMSTELL, it is implemented with the 3D coordinate mapping. Eigenmode analysis for the *H*1 $\boldsymbol{B}$ representation is reported in Ref. [63]. Findings from that research are relevant to the use of *H*1 $\boldsymbol{B}$ in NIMSTELL computations.

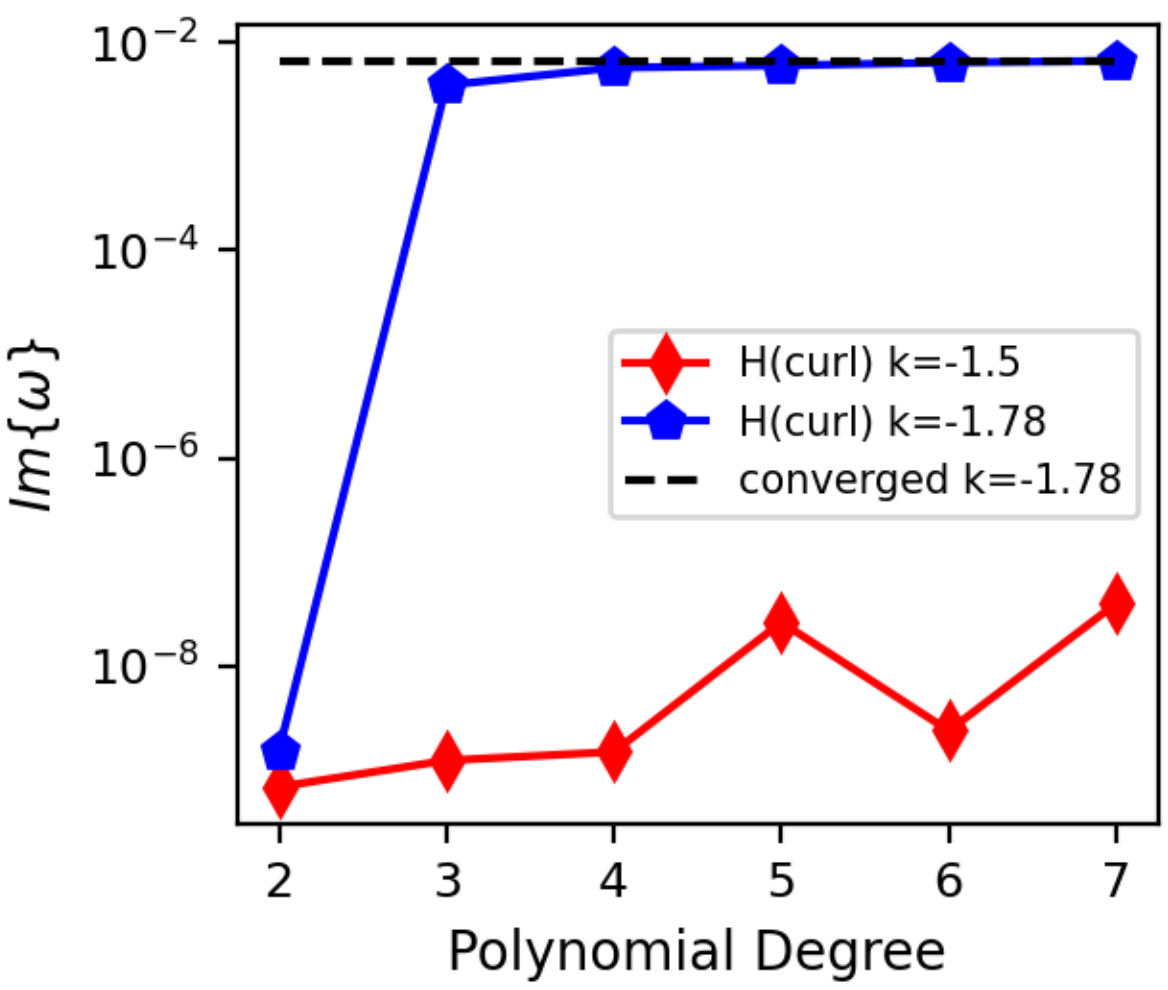

Figure 17. Convergence results with $p$-refinement for 40-element computations of the stable and unstable wavenumbers considered previously. The horizontal axis shows the degree of the expansions used for $A_r$, $p$, and $\boldsymbol{V}$, and the $\boldsymbol{H}$(curl) elements have $rA_\theta$ and $A_z$ of one degree larger.

**Appendix B. Numerical continuity of $\nabla \times \boldsymbol{A}$ across element interfaces**

The $H1$ continuity of the mapping and the shared coefficients for the GLL expansions imply that the representation of $\boldsymbol{A}$ in Eq. (11) has continuous tangential components and that its curl has continuous normal components [52]. Given the non-standard NIMSTELL combination of 2D $\boldsymbol{H}$(curl) elements with a 1D Fourier expansion, we prove these properties in this appendix.

Each element has its own mapping, *i.e.*, its own transformation between element coordinates and physical coordinates. Nonetheless, the continuity across elements implies that the tangent basis vectors that run along the sides of a 2D element, $\partial\boldsymbol{R}/\partial\eta$ and $\partial\boldsymbol{R}/\partial\zeta$ at $\xi = 0, 1$ for example, are continuous. The continuity of the tangential components of $\boldsymbol{A}$ then follows from $\nabla u_i \cdot \partial\boldsymbol{R}/\partial u_j = \delta_{ij}$ and the shared coefficients of the expansion. For example,

$$\boldsymbol{A}_e \cdot \frac{\partial \boldsymbol{R}}{\partial \eta}\bigg|_{\xi=1} = \sum_{k',l'} A_{e,\eta_{k',l'}}(\zeta)\lambda_{k'}(1)\varsigma_{l'}(\eta)\,, \tag{B1}$$

where the dot product selects coefficients that are shared with the adjacent element, regardless of whether the two sets of element coordinates have the same orientation in physical space.

For the continuity of the normal component of $\nabla \times \boldsymbol{A}$ along element interfaces, again consider the representation at $\xi = 1$. Since the Jacobian is

$$J = \frac{\partial \boldsymbol{R}}{\partial \xi} \cdot \frac{\partial \boldsymbol{R}}{\partial \eta} \times \frac{\partial \boldsymbol{R}}{\partial \zeta} = \frac{1}{\nabla\xi \cdot \nabla\eta \times \nabla\zeta}\,, \tag{B2}$$

and $\nabla\xi$ is parallel to $\frac{\partial \boldsymbol{R}}{\partial \eta} \times \frac{\partial \boldsymbol{R}}{\partial \zeta}$, the unit normal along this surface is

$$\hat{\boldsymbol{n}}|_{\xi=1} = \frac{1}{|\nabla\xi|}\nabla\xi\bigg|_{\xi=1} = \frac{J}{\left|\frac{\partial \boldsymbol{R}}{\partial \eta} \times \frac{\partial \boldsymbol{R}}{\partial \zeta}\right|}\nabla\xi\Bigg|_{\xi=1}\,. \tag{B3}$$

The normal component of $\nabla \times \boldsymbol{A}_e$ has contributions from the second and third sums in Eq. (11):

$$\nabla \times \boldsymbol{A}_e \cdot \hat{\boldsymbol{n}}|_{\xi=1} = \left[\sum_{k'',l''} A_{e,\zeta_{k'',l''}}(\zeta)\lambda_{k''}(1)\dot{\lambda}_{l''}(\eta) - \sum_{k',l'} \dot{A}_{e,\eta_{k',l'}}(\zeta)\lambda_{k'}(1)\varsigma_{l'}(\eta)\right] \frac{1}{\left|\frac{\partial \boldsymbol{R}}{\partial \eta}\times\frac{\partial \boldsymbol{R}}{\partial \zeta}\right|}\Bigg|_{\xi=1}, \text{(B4)}$$

where the dot above a function represents its derivative. The only differentiation in (A4) is along the 1D basis functions that lie within the interface, which are the same for adjacent elements. Also, the coefficients are those for an endpoint of the GLL $\lambda_k(\xi)$ polynomials, which are shared for adjacent elements. Thus, apart from a minus sign for outward normal vectors relative to each element, the above expression matches that of the adjacent element. The continuity properties for tangential components of $\boldsymbol{A}$ and for the normal component of $\nabla \times \boldsymbol{A}$ hold regardless of the degree of each polynomial expansion in Eq. (11), and the two terms in (B4) are of the same degree in $\eta$ when the GLL $\lambda_{l''}(\eta)$ is of one degree larger than the GL $\varsigma_{l'}(\eta)$.